# Knowing Is Not Enough: Information Retrievability as a Precondition to Effective LLM Oversight

**Xinyu Fu**
Robinson College of Business
Georgia State University
55 Park Place NE, Atlanta, GA 30303, USA
Email: xinyufu@gsu.edu

**Narayan Ramasubbu**
School of Business
University of Pittsburgh
3950 Roberto and Vera Clemente Dr, Pittsburgh, PA 15260, USA
Email: narayanr@pitt.edu

**Dennis F. Galletta**
School of Business
University of Pittsburgh
3950 Roberto and Vera Clemente Dr, Pittsburgh, PA 15260, USA
Email: galletta@pitt.edu

**Abstract**

Large language models (LLMs) are increasingly embedded in organizational work, yet their errors often pass human review. Prior research locates such failures in users' capability to review LLM output or their engagement in doing so. We develop an alternative, retrieval-based account of human oversight and posit that error detection is more effective when oversight-relevant information is accessible to users at the moment of review. Across two randomized lab-in-the-field experiments with 640 customer-facing employees, we show that self-generated explanations improve error detection and strengthen recall of verification-relevant reasoning, while cues that reactivate such reasoning help sustain detection under repeated LLM use. Theoretically, we identify information retrievability as a distinct precondition for effective oversight and specify generative encoding and cue-supported reactivation as mechanisms that build and sustain it. Practically, lightweight onboarding self-explanations and daily retrieval cues can make human oversight more resilient as LLM use becomes routine.

## Introduction

Organizations are increasingly adopting large language models (LLMs) to augment knowledge work, such as drafting client communications, answering support requests, and preparing analyses, while retaining human reviewers as a safeguard against the models' errors [5,24]. This person-in-the-loop design relies on employees to provide oversight and catch consequential mistakes before they reach customers or undermine downstream decisions. However, errors frequently persist through review, even when reviewers are aware of LLM fallibility and prominent system warnings are displayed. For example, lawyers have filed court briefs citing cases that a chatbot fabricated [7]. One common explanation is that fluent, confident, and largely correct LLM output invites acceptance rather than detailed scrutiny. Even a reviewer who was recently trained on a type of LLM error, or has personally encountered it, may still miss the next instance [18,22]. This is the puzzle that motivates our study: why do users miss errors in LLM-generated output even when they have both the ability and the motivation to catch them?

Extant explanations of LLM oversight failures fall into two broad categories. One locates the failure in capability: users may lack the AI literacy, domain expertise, or mental models needed to review the output [2,15,38]. The other locates it in engagement: even capable users may under-scrutinize a seemingly reliable system, deferring through automation bias, complacency, or miscalibrated trust rather than investing the effort that verification requires [6,17,32]. Inspired by these two schools of thought, researchers and practitioners have developed interventions ranging from provisioning system-provided explanations and performance information to incorporating reflection prompts and cognitive forcing functions [1,3,6,13,42]. These approaches address what users know or how much effort they direct toward review. They leave open a further possibility: a user may possess the oversight-relevant information (i.e., awareness that the LLM system can err, knowledge of recurring error patterns, and strategies for verification) and be motivated to review the output yet fail to retrieve that information when reviewing a particular output because the situation does not cue its retrieval.

Consider the Moses illusion for illustration [8]. Readers who know that Noah, not Moses, is associated with the ark often miss the anomaly in the question "How many animals did Moses take on the ark?" but they answer it correctly when the relevant fact is probed directly. As the example shows, just possessing relevant information does not automatically ensure effective review in a particular context. Similarly, fluent, coherent, and apparently plausible LLM output does not provide sufficient cues for reviewers to draw on the recurring error patterns or relevant verification strategies they have previously learned. The problem is not necessarily that users are not capable or fail to initiate review, but that the information that would make that review effective does not become accessible while the review is underway.

To address this, we develop a retrieval-based account of human oversight grounded in the encoding specificity principle [36]. This principle holds that retrieval depends on the correspondence between the cues available at review and the way information was initially encoded; consequently, acquired information may remain inaccessible in the absence of matching cues [35]. Building on this, we identify retrievability as an important precondition for effective LLM oversight alongside users' capability and engagement.

Viewed through this lens, LLM oversight can be strengthened at two points: how oversight-relevant information is *encoded* and how it is later *retrieved* [11,36]. Encoding concerns how information is first represented. For example, generating an explanation, rather than merely reading one, produces a more elaborated and distinctively organized representation of oversight-relevant information [9]. Retrieval concerns the later moment of use. A brief cue that the user associates with their earlier oversight-relevant reasoning can reactivate that reasoning when a new LLM output is reviewed, even when additional, detailed information is not supplied [31]. Because the self-generated explanation and reasoning were originally formed by working through why a specific output was wrong, reactivating it through a matching cue restores the specific checks and verification strategies that the error called for, along with the recognition that fluent, plausible LLM output can still have mistakes, warranting scrutiny.

We test this retrieval-based account of oversight effectiveness in two randomized lab-in-the-field experiments with employees of a large enterprise-software organization in China. Study 1 ($N = 400$) isolates encoding while holding constant the oversight-relevant information: participants either generated their own explanation of a set of LLM errors before receiving a standardized explanation or received that explanation without first generating their own. Self-explanation produced more elaborated recall and approximately a 10-percentage-point improvement in subsequent error detection. Study 2 ($N = 240$), conducted over a baseline day and seven subsequent workdays, holds encoding constant and randomizes retrieval support. All participants generated an explanation at baseline. They then customized a daily message that appeared during subsequent use: in one condition the message carried a brief, self-chosen cue tied to their earlier reasoning, and in the other it carried an equally personalized but verification-neutral phrase. Error detection declined significantly more slowly when the daily message carried the participant's cue. Together with the cue-evocation and cue–episode association measures, this trajectory difference is consistent with the cue reactivating the participant's earlier reasoning rather than functioning only as a personalized message.

This research makes three contributions. First, we identify the retrievability of oversight-relevant information as a distinct and complementary determinant of effective LLM oversight.

Oversight can fail even when users possess the capability and motivation to detect errors because the review context may provide no cue that matches how their verification-relevant reasoning was encoded. The reasoning can remain intact yet inaccessible at the moment it is needed. Second, our results indicate that oversight support benefits from active encoding of verification-relevant reasoning, not exposure to oversight-relevant information alone. Generating an explanation creates an elaborated and personally organized representation of verification-relevant reasoning that remains more accessible during later review than equivalent content received passively. Third, our results indicate that sustaining oversight need not require continually supplying new warnings or explanations. As error detection declines over repeated use, a cue that reactivates reasoning the user has already formed can slow that decline without repeating the original explanation or providing task-specific corrective information. Reactivation offers an alternative to repeated warnings or information provision.

**Literature Review**

Human oversight of an LLM refers to the review of its output before it is accepted, forwarded, or acted upon. Oversight can take several forms, including examining specific claims, consulting relevant evidence, and deciding whether to accept, revise, or reject the output. In this study, we focus on error detection as a central manifestation of effective oversight. Such oversight does not require generalized skepticism or blanket rejection of LLM output. It requires calibrated review, accepting accurate content and intervening when it is erroneous [17,21]. LLM errors often survive review even when users have already learned that the system can make mistakes [18,22]. Among user-centered accounts, prior research offers two explanations for this oversight failure. A capability-based explanation focuses on whether users possess accurate and usable evaluative resources needed to assess the output. In contrast, an engagement-based explanation considers whether users initiate and sustain the attention and effort required to apply those resources. Both explanations are established in conventional decision support settings, where systems provide bounded recommendations, scores, or diagnoses, and increasingly inform research on generative artificial intelligence (GenAI) applications, where users review extensive, multi-claim content. In Online Appendix A, we provide a detailed summary of the literatures on capability- and engagement-based explanations.

***The User-Capability Explanation***

The capability-based explanation attributes oversight failure to the lack of accurate and usable evaluative resources that a user needs to scrutinize GenAI output [15,18,38]. These resources concern whether users understand what a GenAI system can and cannot do and how much evidentiary weight its output warrants, ranging from objective knowledge of what GenAI systems do to broader AI literacy [38] to system-specific mental models of how a particular system tends to fail [15]. In addition, task-specific resources concern whether users command the domain expertise, reference evidence, and verification procedures needed to adjudicate a particular claim [2,30,39]. Corresponding interventions supply or correct these resources through AI-literacy training, disclosures of system limitations, performance and uncertainty information, and system-provided explanations, with benefits that are real but uneven [3,13,42,43]. We refer to the content within these resources that bears on reviewing a given output as oversight-relevant information: awareness that a system can err, its competence boundaries and recurring error patterns, task knowledge relevant to verifying a claim, and strategies for verification. Capability-based accounts explain whether users possess adequate oversight-relevant information; they do not explain whether users undertake and sustain the evaluative activity needed to apply it, the question the engagement-based account takes up next.

### *The Engagement Explanation*

The second explanation locates oversight failure in engagement: whether users initiate, allocate, and sustain the attention and evaluative effort needed to scrutinize AI output and regulate reliance on the system [28,32]. Users may substitute the system's output for their own information processing, the automation bias that produces both omission and commission errors [22,25,32], and they may fail to sustain scrutiny over repeated interaction, because monitoring a usually reliable system offers little immediate return and reliance drifts away from the system's actual reliability [17,21,28]. Corresponding interventions interrupt default acceptance through cognitive forcing functions, reflection prompts, and displays of confidence or performance information [1,6,42]. GenAI systems intensify both of these engagement problems as they produce fluent, mostly correct drafts that make exhaustive verification costly and routine acceptance attractive [7,37]. Engagement-based accounts explain whether users initiate and sustain evaluative activity; together with capability-based accounts, they describe whether users possess relevant resources and actively review the output. Neither explains whether the particular oversight-relevant information required for a given output becomes accessible during that review. This gap is explored in the next section.

### *Retrieval of Oversight-Relevant Information*

Capability- and engagement-based interventions address two complementary prerequisites of effective oversight. Capability concerns whether users possess adequate evaluative resources, whereas engagement concerns whether they initiate and sustain the activity needed to apply them. We identify retrievability as whether the oversight-relevant information required for verifying a particular GenAI output becomes accessible at the moment it is reviewed. Extant capability- and engagement-based accounts generally leave this condition implicit. They assume that once relevant oversight information has been acquired and users are actively reviewing an output, that information will be available for use. Yet relevant information may be retained but temporarily inaccessible, and users may actively scrutinize an output without bringing the pertinent oversight-relevant information or reasoning to mind. Knowing that verification is warranted is therefore not equivalent to having the specific information needed for effective verification accessible in that moment. Table 1 summarizes these three user-side conditions, and Online Appendix A reviews representative prior work across them.

**Table 1. Three Complementary User-Side Conditions for Effective Oversight of LLM Output**

| | Capability | Engagement | Retrievability (This Study) |
|---|---|---|---|
| What the account explains | Whether users possess the information and procedures needed to review LLM output | Whether users undertake and sustain the evaluative activity needed to apply those resources | Whether relevant possessed information becomes accessible and guides a particular review |
| Primary mechanisms | AI-specific resources; task-specific resources | Initiating and allocating independent scrutiny; sustaining scrutiny and regulating reliance | Encoding that supports later accessibility; cue-supported reactivation at the point of use |
| Locus of oversight failure | Relevant resources are absent, incomplete, inaccurate, or insufficient for distinguishing correct from erroneous content | Independent scrutiny is not initiated or sustained, or excessive reliance displaces verification | Oversight-relevant information is retained but not retrieved when a particular output is reviewed |
| Representative interventions | AI-literacy training; disclosures of system limitations; performance and uncertainty information; system-provided explanations; domain and verification support | Generic warnings; cognitive forcing functions; reflection prompts; interface friction | Generative encoding of verification-relevant reasoning; reactivation through associated retrieval cues |
| Representative work | [2,15,16,18,30,38,39] | [1,6,17,20,21,22,25,28,32,42] | — |

*Note.* Effective oversight may also depend on artifact and organizational conditions, such as whether the system exposes diagnostic evidence and whether users have the authority to intervene. We focus on the user-side conditions in this paper.

Leaving retrievability implicit creates a theoretical blind spot in prevailing accounts of oversight. Many interventions provide information through training, explanations, or warnings, assuming that the information will later be accessible when users review new output. When that information is retained but not retrieved, an adequate intervention may appear ineffective. A weak or null effect may arise from a retrieval failure rather than from deficiencies in the information supplied or in users' willingness to review the output. Ignoring this distinction can lead research to misdiagnose why oversight support succeeds or fails. Although research on professional judgment recognizes that verification depends partly on how accessible and well-organized relevant information is, not only on whether it is possessed [14], it has not developed retrievability as a distinct, designable condition of human oversight of LLM output. In particular, prior oversight research has not explained how recently encoded oversight-relevant information can be retained yet fail to become accessible when a new LLM output is reviewed, or how design can shape that accessibility. Our studies examine a consequential and designable instance of this broader bottleneck: whether recently acquired, task-specific oversight reasoning becomes accessible during subsequent review. We develop a retrieval-based account of how the encoding of oversight-relevant information and the cues available during review shape its retrievability.

## Theory and Hypotheses

In developing our retrieval-based account of LLM oversight, we first distinguish the encoding, retrievability, and retrieval of oversight-relevant information and establish two relationships between them. Initial encoding shapes how retrievable the information becomes, whereas cues available at the point of review support its retrieval. We then apply these relationships to generative encoding and cue-supported reactivation and develop theoretical predictions for each.

### *A Retrieval-Based Account of Oversight*

Exposure to oversight-relevant information does not ensure its later retrieval. For such information to guide verification and error detection, it must first be encoded and retained and then retrieved when a subsequent LLM output is reviewed. Information can remain available in memory without being accessible at the moment it is needed, a long-standing distinction between availability and accessibility [12,35]. We use retrievability to refer to the potential for encoded oversight-relevant information to become accessible later, and retrieval to refer to the act of accessing it during review. Only information that is retrieved can directly guide verification of the current LLM output.

How information is initially encoded shapes its later retrievability. Encoding is not a uniform consequence of exposure. The same content can be represented with different degrees of elaboration, organization, and distinctiveness. More elaborated and organized representations generally better support later access than surface familiarity alone [11]. Later access is also stronger when the operations performed during encoding resemble those required at the point of use, because the resulting representation is better matched to the later task [27]. Generative activity can therefore improve retrievability by requiring users to construct and organize information rather than merely receive it, consistent with the established advantage of generated over passively read information in later retrieval [9].

Retrievability alone, however, does not guarantee retrieval. Whether encoded information becomes accessible also depends on the cues present when it is later needed. Under the encoding specificity principle, a cue supports retrieval to the extent that it reinstates features associated with the original encoding episode [36]. Thus, the usefulness of a cue depends on its association with the target representation. A salient reminder may attract attention without reinstating the particular information required for verification, whereas an associated cue can reconnect the current review context to a previously encoded representation [31]. A retrieval-based account therefore treats later access as jointly shaped by the representation formed during initial encoding and the cues available at the point of review. These two points provide complementary design levers. Generative encoding can produce a representation with greater retrievability, whereas associated cues can support its later retrieval. Together, these pathways explain why users who possess the same information and actively review the same output may differ in whether that information guides verification and error detection. The following sections apply these two relationships to oversight of LLM-generated output.

### *Generative Encoding of Oversight-Relevant Information*

Self-explanation instantiates the generative-encoding pathway in LLM oversight. In self-explanation, a learner articulates why an example is correct or incorrect, thereby elaborating and organizing the relevant information in relation to their own understanding. Prior research shows that self-explanation improves comprehension and transfer by engaging users in constructing rather than passively receiving explanatory content [4,9,10].

Consider two users who receive the same accurate explanation of an LLM error but differ in the activity that precedes it. A user who only reads the explanation processes content already organized by someone else. A user who first generates an explanation must identify the problematic element, connect it to the relevant reference information, and articulate why the output is incorrect. Equivalent explanatory information therefore might not produce equivalent encoding [11,12]. We refer to the elaborated, self-organized representation produced through this activity as verification-relevant reasoning, a specific form of oversight-relevant information.

Generative encoding of verification-relevant reasoning should make that reasoning more retrievable during subsequent review than receiving the same explanatory information passively because generation engages operations resembling those later required for verification [27]. When users later review new LLM-generated drafts, more retrievable verification-relevant reasoning is more likely to become accessible and guide them in identifying relevant error patterns, recalling verification strategies, and comparing claims against reference materials. Once accessible, that reasoning can support the detection of erroneous content. The account does not require verbatim or conscious recall of the original explanation. It requires only that the encoded reasoning becomes accessible enough to guide verification. Generative encoding of verification-relevant reasoning should therefore improve subsequent error detection.

> ***Hypothesis 1 (H1).*** *Users who first generate their own explanation of an LLM error will detect more errors in subsequent LLM output than users who receive the same standardized explanation without first generating one.*

***Cue-Supported Reactivation Under Repeated Use***

A cue associated with a user's prior reasoning instantiates the retrieval-support pathway. Because retrieval is cue-dependent, the later appearance of a cue associated with encoded verification-relevant reasoning can reinstate that representation without restating the original explanation or providing task-specific corrective information. Retrieval support therefore depends on the association between the cue and the encoded reasoning, not merely on the cue's salience or informational content. Consider an otherwise unrelated private keyword such as "pineapple" that becomes linked to a particular prior episode. To an outsider, the word conveys nothing about the reasoning; to the person who formed the association, it can later reactivate the reasoning from that episode. A cue drawn from a user's own earlier explanation can therefore reconnect the current review context to that user's prior reasoning more effectively than a generic reminder of comparable prominence [31,36]. Online Appendix B provides a step-by-step illustration of this mechanism.

Retrieval support becomes critical under repeated use. Each new output creates another occasion on which previously encoded reasoning must be retrieved. As time and intervening tasks separate the original encoding episode from later review, the reasoning may remain retained yet fail to become accessible when the review context lacks appropriate retrieval cues, consistent with established accounts of forgetting and retrieval failure [35,36,40]. A cue associated with the user's earlier explanation can reinstate that reasoning at the point of review. Repeated retrieval can also strengthen subsequent access rather than merely provide another exposure to the information [26].

Repeated use may additionally foster routine, habituation, and complacency, which can reduce the scrutiny users direct toward system output [23,28,37]. These engagement processes can coexist with declining accessibility. What distinguishes the retrieval-based account from a generic increase in vigilance is therefore association rather than salience. Habituation or complacency may predict that any prominent reminder temporarily increases scrutiny. They do not predict a specific advantage for a cue associated with the user's previously encoded reasoning over a comparably presented personalized message generated without that association.

Because cue-supported reactivation operates against an ongoing loss of accessibility, it should attenuate rather than eliminate the decline in error detection. A cue associated with prior verification-relevant reasoning should preserve access, and thereby sustain error detection more effectively across repeated use than a comparably presented personalized message generated without that association. The distinctive prediction is therefore longitudinal. Association with prior reasoning should slow the decline in error detection.

> ***Hypothesis 2 (H2).*** *During repeated LLM-assisted work, error detection will decline more slowly among users who receive a cue associated with their prior verification-relevant reasoning than among users who receive a comparably presented personalized message generated without that cue association.*

Figure 1 summarizes the resulting model, linking encoding, retrieval, verification, and error detection. Table 2 defines these constructs and their study-level instantiations.

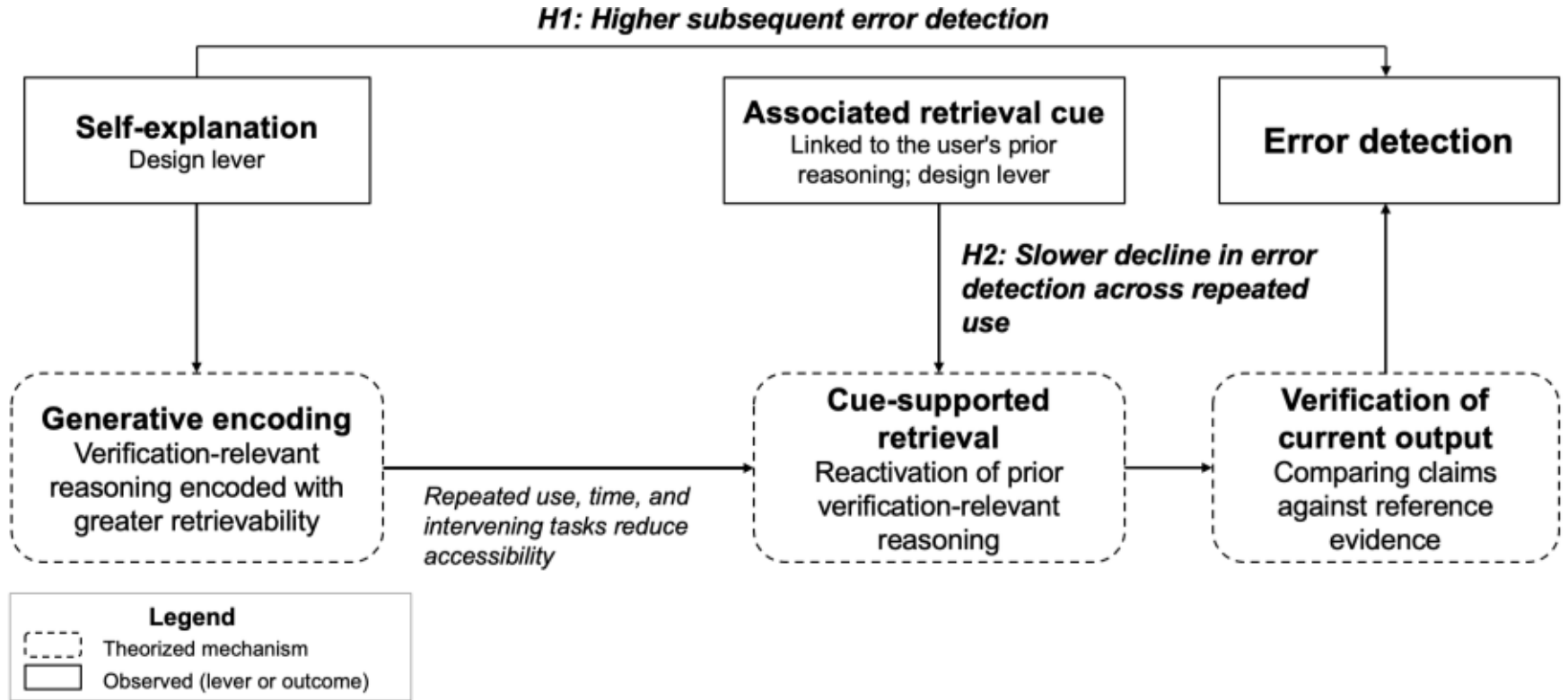

**Figure 1. A Retrieval-Based Model of Human Oversight in LLM-Assisted Work**

**Table 2. Core Constructs in the Retrieval-Based Model of Human Oversight**

| Construct | Definition | Instantiation in the present studies |
|---|---|---|
| Oversight-relevant information | Information a reviewer can draw on to check machine-generated output, comprising awareness that a system can err, its competence boundaries and recurring error patterns, task knowledge relevant to verifying a claim, and strategies for verification. | Awareness that LLM drafts can err (provided to all participants), instantiated more specifically as verification-relevant reasoning about the illustrative errors (next row). |
| Verification-relevant reasoning | The specific encoded reasoning about why a given output may be wrong and how to confirm it; a study-level instantiation of oversight-relevant information. | A reviewer's articulated account of why a stated figure is implausible and how to verify it against reference materials. |
| Encoding | How oversight-relevant information is initially processed and stored; the antecedent manipulated in Study 1. | Generating one's own explanation of an error versus reading a system-provided explanation. |
| Generative encoding | Encoding through self-generation of an explanation; an instantiation of encoding, not a separate mechanism. | Self-explanation: Writing, in one's own words, why each illustrative LLM error is wrong. |
| Retrievability | The potential for encoded oversight-relevant information to become accessible at the moment of review; addressed experimentally through its two determinants, encoding (Study 1) and retrieval support (Study 2). | Whether earlier verification-relevant reasoning comes to mind when a new LLM output is reviewed. |
| Retrieval (reactivation) | The act of accessing or reactivating previously encoded information at the point of review. | A cue bringing the earlier self-explanation to mind during later review. |

## Research Setting and Overview

### *Organizational Setting and LLM-Assisted Workflow*

We conducted two lab-in-the-field experiments in collaboration with a large enterprise-software organization in China operating in the enterprise resource planning and digital management industry. The focal business unit consists of customer-facing knowledge workers, primarily in sales and pre-sales roles. These employees routinely handle a high volume of customer inquiries, retrieve case and project information, configure products and solutions, and prepare customer communications and downstream recommendations. In 2025, several months before data collection, the organization deployed an internally developed LLM assistant to support these activities, including retrieving internal knowledge and drafting responses to customer inquiries. Employees were expected to verify the generated content against internal documentation before communicating it to customers or using it in downstream decisions. Error detection and verification were therefore embedded within a broader, high-throughput workflow rather than performed as isolated or primary tasks.

In December 2025, when the studies were conducted, the assistant supported 20,878 employee–LLM query interactions from 781 active employees in the focal business unit (median 26 per active user, range 13–68), with a stable daily volume of roughly 670 interactions. These queries spanned customer case and project retrieval, HR and compensation processes, industry-solution design, product configuration, technical integration, and the use of AI tools. Together, these patterns characterize a routine, high-volume, multi-domain LLM-assisted workflow.

***Participants and Experiment Procedures***

Participants were drawn from the same business unit but constituted separate, non-overlapping samples; no employee took part in both studies. Data for both studies were collected in December 2025. All procedures were approved by the authors' institutional review board. Participation was voluntary, and all data were anonymized before analysis. Across both studies, participants completed information-verification tasks that reproduced a central evaluative demand of the organization's routine customer-facing workflow: determining whether a plausible LLM-generated draft could be relied upon before use. Participants evaluated fluent, professional-looking LLM-generated drafts using searchable authoritative reference materials adapted from internal knowledge articles, product documentation, and policy texts. Study 1 presented 10 email-verification tasks in a single session, whereas Study 2 interspersed test emails with participants' routine work across repeated workdays. The experimental drafts were fixed before data collection rather than generated in real time. To construct the materials, the researchers identified recurring error patterns in actual outputs produced by the organization's LLM assistant and manually embedded representative instances of these errors into otherwise accurate drafts. Each Study 1 test email contained exactly one predefined, verifiable error, whereas each Study 2 test email could contain multiple predefined errors. In both studies, the errors were embedded among otherwise correct statements and were drawn from three recurring categories of enterprise-support inaccuracy: applicability or condition errors, constraint or exception errors, and procedural errors (Table 3). The reference materials were sufficient to resolve the embedded errors but required participants to actively consult them and compare them against the LLM-generated drafts. Successful performance therefore required selective verification rather than accepting or rejecting the drafts as a whole.

The two studies share the same three error categories, searchable reference materials and review interface, and ground-truth resolution with blind double coding (see the Measurements section); scoring and aggregation follow each study's task structure (see the study-specific measures for Study 1 and Study 2). They differ in the theoretically targeted intervention: Study 1 varies how oversight-relevant information is initially encoded within a single session, whereas Study 2 holds generative encoding constant and varies subsequent retrieval support across seven workdays. The procedures are summarized in Figure 2.

**Table 3. Error Categories Used in Both Studies, with Illustrative Examples**

| Error category | Definition | Illustrative example | Verification check |
|---|---|---|---|
| Applicability or condition error | Otherwise plausible content misapplied to a case for which it does not hold because the version, role, region, or other applicability criterion differs. | A draft answer recommends a delivery and configuration approach that applies only to the premium private-cloud edition of the platform, although the customer's inquiry concerns the standard cloud edition. | Check whether the recommendation applies to the specific case, including the product version, user role, region, and other applicability criteria. |
| Constraint or exception error | Applicable content presented without a qualification, threshold, limitation, exception, or exclusion needed to delimit when or to what extent it holds. | A draft answer recommends a performance-management module without noting that it is designed for organizations above a documented employee-count threshold. | Check whether the claim requires a threshold, qualification, limitation, exception, or exclusion. |
| Procedural error | Incorrect or incomplete steps, ordering, approvals, prerequisites, or required actions in a process. | A draft answer describes the payroll disbursement process but instructs the user to submit the payment file through the bank interface before the required | Check prerequisites, sequence, approvals, and required actions against the documented process. |

approval workflow has been completed.

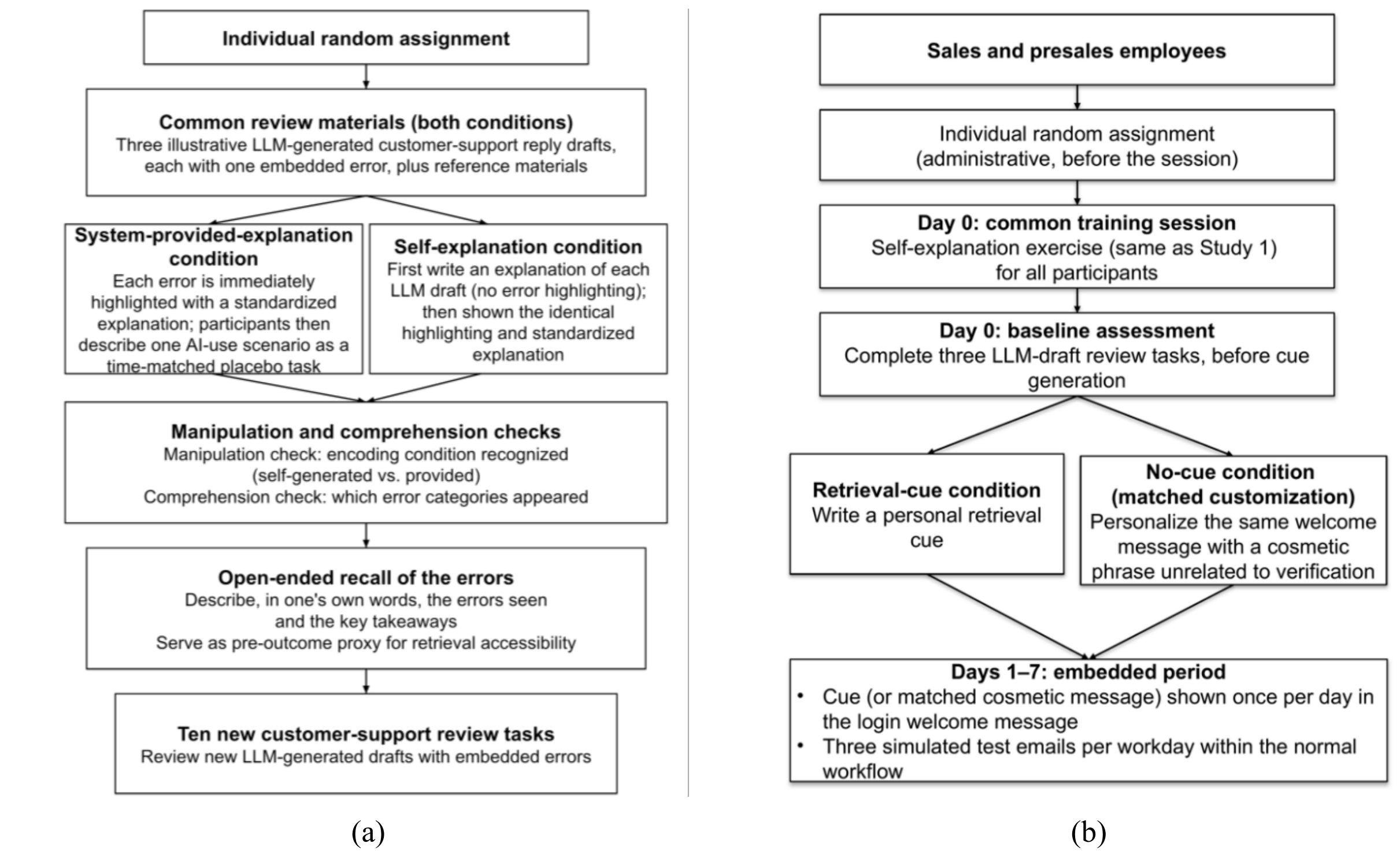

**Figure 2. Experimental Procedures of Study 1 (a) and Study 2 (b)**

***Measurements***

The primary dependent variable in both studies was behavioral error detection, operationalized as participants' success in resolving predefined errors embedded in otherwise plausible LLM-generated drafts. The unit of coding was each predefined error, scored against a predefined ground truth. The two studies apply this common procedure at different granularities because their task structures differ: in Study 1 with a single error per email, partial credit captures the degree to which that error was resolved, whereas in Study 2 with multiple errors per email, binary per-error coding combined with the proportion corrected yields a comparable graded rate. In both studies, the dependent variable is an error-detection rate on a 0–1 scale, analyzed at the participant level in Study 1 and at the participant-day level in Study 2. The measures sections for Study 1 and Study 2 give the scoring scheme used in each study and describe how the resulting error-level scores were aggregated. Before data collection, the researchers specified all embedded errors, acceptable resolutions, and scoring rules based on authoritative organizational documentation. Participants' final responses were then evaluated against this predefined ground truth using the prespecified scoring criteria. Two coders blind to experimental condition independently scored each response against the prespecified resolution criteria, yielding high interrater agreement (Study 1: Cohen's $\kappa = .92$; Study 2: $\kappa = .89$). Discrepancies were resolved through joint review against the prespecified criteria, with any unresolved cases adjudicated by a third coder blind to experimental condition.

Because higher scores could reflect a general tendency to distrust and alter LLM output rather than accurate error detection, we also recorded false positives, coded at the level of individual statements: each test email contained 10–13 verifiable correct statements, and a false positive was recorded when a participant flagged, removed, or modified a correct statement. A participant's false-positive rate is the mean of the email-level rates (the share of correct statements altered in each email), not a pooled ratio across emails; the daily false-positive rate in Study 2 follows the same statement-level definition. Because error detection is coded at the level of predefined errors and false positives at the level of correct statements, the net-detection index used in the robustness analyses contrasts two rates defined on different units and is interpreted as a summary index rather than a single-scale rate.

Both studies additionally recorded auxiliary indicators of verification activity, including time spent reviewing the LLM-generated draft, consultation of reference materials, and the extent of changes made before submission. These measures were used to assess whether the manipulations affected verification behavior more broadly. Finally, both studies included three participant-level covariates aligned with the capability and engagement prerequisites developed in the literature review. Objective AI knowledge was assessed using two multiple-choice items.

Our primary covariate is a binary indicator of whether the participant correctly identified the breadth of AI application domains on the first item. As a robustness check, we also use a 0–2 score combining both knowledge items; treatment estimates are unchanged (Online Appendix D). Domain capability is measured as the natural logarithm of one plus the participant's sales-deal amount for the focal product line during the month preceding the experiment, with no recorded sales coded as zero. Because B2B deal closures are sparse over short horizons and longer historical windows are less comparable given mobility across sales roles, most participants recorded no sales during this window (94% in Study 1 and 92% in Study 2). Results are unchanged using a binary indicator of any recorded sales or longer-horizon measures of sales experience and performance (Online Appendix D). Oversight orientation captures the engagement-related disposition to scrutinize rather than accept AI output; it is a binary indicator coded 1 when the participant endorsed either of two reasons for not accepting AI-assisted selling, namely that the accuracy of AI work cannot be guaranteed for the company's complex product system or that the quality, validity, and reliability of AI models are unclear.

Because these covariates are measured independently of the experimental tasks (i.e., prior to the treatment stimuli exposure), they support randomization checks and covariate adjustment without absorbing treatment-related variance; verification behaviors recorded during the tasks (e.g., verification time) are post-treatment outcomes and are analyzed as parallel dependent variables rather than covariates. All study materials and measures were administered in Chinese, the participants' working language; Online Appendix C presents the original Chinese wording and illustrative English translations.

***Two-Study Mechanism-Testing Strategy***

The logic underlying our hypotheses is that the interventions will improve error detection by increasing the retrievability of oversight-relevant information. Testing this proposed mechanism therefore requires evidence that the interventions operate on the retrieval process specified by the theory. A conventional mediation analysis, however, is not feasible because retrievability is a latent, context-dependent cognitive property for which we lack a valid measure that can be situated between the intervention and the outcome. In such settings, a coordinated experimental causal-chain strategy can provide process evidence through three coordinated steps [33]. First, the proposed mechanism is decomposed into theoretically distinct points at which it can be influenced. Second, experiments intervene separately on those points while holding other parts of the process as constant as possible. Third, the resulting pattern of proximal cognitive responses and downstream outcomes is examined for consistency with the proposed process. The mechanism is therefore evaluated through the coordinated consequences of theoretically targeted interventions rather than solely through the statistical association between a measured mediator and an outcome. This strategy is well established in organizational and information systems research, both for manipulating theoretically distinct components of a proposed process [34] and for probing a cognitive process that resists direct measurement by varying the conditions under which it should operate [41]. It is especially appropriate for our analysis because a measured proximal indicator would remain a post-treatment variable whose association with the outcome can be confounded even under random assignment [29].

We implement this strategy by targeting the two points at which retrievability can be shaped: initial encoding and subsequent retrieval support. Study 1 intervenes on initial encoding by comparing generative explanation with an equivalent system-provided explanation while holding later retrieval support constant. It examines whether generative encoding improves both pre-outcome recall and subsequent error detection (H1). Study 2 holds initial generative encoding constant and intervenes on subsequent retrieval support by varying whether participants receive a cue associated with their prior reasoning. It examines whether the cue evokes that prior reasoning and slows the decline in error detection across repeated use (H2). Thus, the two studies do not simply replicate the same treatment effect. They test complementary links in the proposed process: whether oversight-relevant information is encoded in a more retrievable form and whether later access to it can be sustained through retrieval support.

The proximal measures, pre-outcome recall in Study 1 and cue evocation in Study 2, provide evidence that each intervention affected the intended part of the process. They are treated as manipulation-linked process indicators rather than direct measures of retrievability or formal statistical mediators. Taken together, the coordinated interventions, proximal responses, and behavioral outcomes provide convergent evidence for the retrieval mechanism.

## Study 1: Generative Encoding and Subsequent Error Detection

### *Participants and Design*

Participants were recruited through an internal invitation to a scheduled workshop on AI-assisted work practices. All employees in the focal office who met the role criteria described above were invited to participate during work hours; no additional eligibility criteria were imposed. A total of 407 employees completed Study 1 and were randomly assigned to either the system-provided-explanation condition ($n = 204$) or the self-explanation condition ($n = 203$). Seven participants failed prespecified attention checks (instructed-response items embedded in the study surveys) and were excluded before analysis (four and three, respectively), resulting in a final analytic sample of 400 participants, with 200 in each condition. A sensitivity analysis indicated that the analytic sample provided 80% power ($\alpha = .05$, two-tailed) to detect a between-condition difference of Cohen's $d = 0.28$ or larger.

### *Experimental Procedure*

Figure 2a summarizes the procedure of Study 1, which had two phases: an encoding phase, in which participants formed their understanding of the errors, and a subsequent unaided review phase, in which no explanation of any kind remained available. At the outset, participants were informed that the LLM assistant could be helpful but that they remained responsible for verifying its output. In the encoding phase, both control and treatment groups were allotted the same 10-minute training period and first reviewed the same three illustrative LLM-generated responses and their accompanying reference materials. In the system-provided-explanation condition, the interface immediately highlighted the erroneous statement in each illustrative response and displayed a standardized explanation based on the reference materials. During the training period, participants in this condition also completed a placebo task in which they described an example of an AI-use scenario. In the self-explanation condition, participants first wrote an explanation of how each response might have been produced and what might be problematic about it, and only then saw the same highlighted error and standardized explanation.

All participants therefore ultimately received identical information identifying and explaining each error, and both conditions were allocated the same total training time. Thus, random assignment of treatment varied the training activity: participants either generated their own verification-relevant reasoning before receiving the standardized feedback or completed the placebo writing task and received that feedback directly. After reviewing the three illustrative examples, all participants completed the three study-specific measures described below. Finally, in the unaided review phase, participants completed 10 new verification tasks in the common task format (described earlier), without access to the illustrative examples, the reference materials that had accompanied them, or any explanatory content (their own self-generated explanations or the standardized explanatory feedback); the reference materials for the new tasks remained available.

***Study-Specific Measures***

In Study 1, the single predefined error in each task was scored 1 (resolved), 0.5 (partly resolved), or 0 (unresolved). The primary error-detection outcome was the participant's average score across the 10 tasks; false-positive and verification measures were aggregated analogously. After completing the illustrative examples and before beginning the outcome tasks, participants completed three study-specific measures. First, a condition-recognition item asked whether they had generated their own explanation before viewing the standardized explanation or had received the standardized explanation directly; this item served as a manipulation check. Second, an error-category-recognition measure presented six categories, three focal categories that had appeared (applicability or condition, constraint or exception, and procedural errors) and three distractor categories that had not, and asked participants to select those they judged to have appeared in the illustrative examples; participants could select any number. Third, an open-ended recall item asked participants to describe the errors they had encountered in the examples and what they had learned from them.

Two coders who were blind to experimental condition independently rated each open-ended response for recall elaboration using a four-point scale. A score of 0 indicated no error mentioned; 1, a generic reference to an error; 2, recall of a specific error type or instance; and 3, articulation of an underlying mechanism, downstream consequence, or generalized oversight lesson extending beyond the specific example. Interrater reliability was high (Cohen's $\kappa = .82$; linear-weighted $\kappa = .86$), and disagreements were resolved through discussion. Study 1 also included a supplementary two-item objective AI-knowledge test, reported in Online Appendix C;

the application-domain item provides the objective AI knowledge covariate noted in the Measurements section.

### *Data Analysis*

We tested H1 by regressing each participant's error-detection rate on a binary self-explanation indicator, first without covariates and then adding objective AI knowledge, domain capability, and oversight orientation. All tests were two-tailed. We report the condition effect as the regression coefficient *b*, which equals the difference in mean error-detection rates, together with its p-value; standard errors appear in the tables. Additionally, we report the false-positive and net-detection estimates alongside the primary results. We present regression estimates for consistent reporting across our two studies, and equivalent ANOVA and ANCOVA statistics are presented in Online Appendix D along with a median-split logistic model and additional robustness checks. We used *R* version 4.6.0 for our analysis.

### *Results*

#### *Randomization and Manipulation Checks*

Table 4 reports participant characteristics by condition, and we did not observe significant differences between the two groups in age, organizational tenure, objective AI knowledge, domain capability, oversight orientation, or gender distribution. Correlations among the behavioral measures and covariates are reported in Online Appendix E, Table E1. Participants also accurately recognized the explanation condition they had experienced. The condition-recognition item was answered consistently with assignment by 94.5% of participants in the self-explanation condition and 91.0% in the system-provided-explanation condition; the recognition rates did not differ significantly across the two groups ($\chi^2(1, N = 400) = 1.82$, $p = .18$).

**Table 4. Descriptive Statistics for Study 1**

| Variable | System-provided explanation | Self-explanation | Range | Variable | System-provided explanation | Self-explanation | Range |
|---|---|---|---|---|---|---|---|
| Age (years) | 29.36 (6.12) | 29.79 (6.60) | 20.00–40.00 | Error-detection rate | 0.68 (0.12) | 0.78 (0.12) | 0.35–1.00 |
| Organizational tenure (years) | 4.06 (2.92) | 4.01 (2.83) | 0.00–14.00 | False-positive rate | 0.12 (0.04) | 0.09 (0.04) | 0.03–0.25 |
| Objective AI knowledge (0/1) | 0.19 (0.39) | 0.16 (0.37) | 0.00–1.00 | Verification time (s) | 134.57 (18.76) | 159.69 (18.70) | 66.80–198.20 |
| Domain capability (log sales) | 0.83 (2.95) | 0.45 (2.09) | 0.00–13.48 | Reference clicks | 2.87 (0.73) | 3.60 (0.81) | 0.82–5.91 |
| Oversight orientation (0/1) | 0.74 (0.44) | 0.77 (0.42) | 0.00–1.00 | Modification extent | 0.27 (0.06) | 0.35 (0.06) | 0.12–0.52 |
| Gender (F / M / NA) | 95 / 92 / 13 | 86 / 106 / 8 | — | | | | |

***Note.*** Cells report means with standard deviations in parentheses (counts for gender). The condition differences in verification time, reference clicks, and modification extent are significant, $t(398) = 13.41$, 9.41, and 13.33, respectively, all $p < .001$. Randomization tests: age $t(398) = -0.68$, tenure $t = 0.17$, objective AI knowledge $t = 0.79$, domain capability $t = 1.49$, oversight orientation $t = -0.70$ (all n.s.); gender $\chi^2(2, N = 400) = 2.63$, $p = .27$.

*Effect of Self-Explanation on Error Detection*

Figure 3 presents error-detection rates by experiment condition. Participants in the self-explanation condition achieved a higher mean error-detection rate than those in the system-provided-explanation condition (*Mean* = 0.78 versus 0.68).

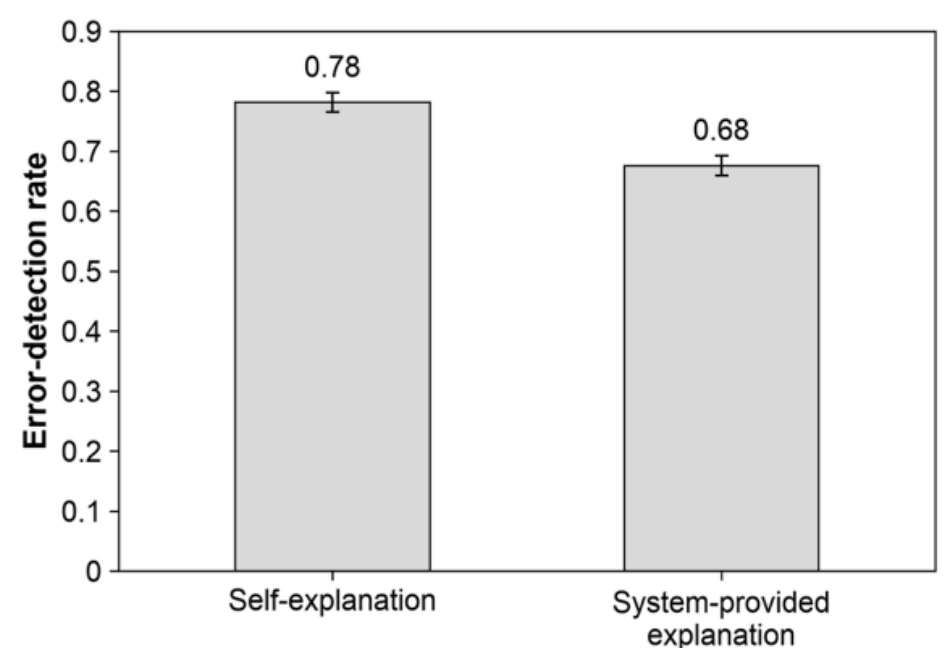

**Figure 3. Study 1 Error Detection by Condition**

*Note.* Error detection is the mean of the 10 per-task scores (each scored 0, 0.5, or 1) per participant; error bars are 95% confidence intervals.

As shown in Table 5, participants in the self-explanation condition detected errors at a significantly higher rate than those in the system-provided-explanation condition ($b = 0.106$, $p < .001$), a difference of about 10.6 percentage points. Because each Study 1 task contained exactly one embedded error scored 0, 0.5, or 1, this difference corresponds to about one additional fully resolved error across the 10 review tasks. The estimate was essentially unchanged when objective AI knowledge, domain capability, and oversight orientation were added as covariates ($b = 0.105$, $p < .001$). The equivalent ANOVA and ANCOVA statistics, reported in Online Appendix D, yield identical conclusions. These results support H1.

**Table 5. Ordinary Least Squares Estimates for Study 1**

| Variable | (1) Unadjusted Model | (2) Full Model | (3) False-Positive Rate | (4) Net Detection |
|---|---|---|---|---|
| Self-explanation | 0.106***<br>(0.012) | 0.105***<br>(0.012) | −0.030***<br>(0.004) | 0.136***<br>(0.013) |
| Objective AI knowledge | | 0.011<br>(0.016) | −0.003<br>(0.005) | 0.014<br>(0.017) |
| Domain capability (log sales) | | −0.001<br>(0.002) | −0.001<br>(0.001) | $0.187 \times 10^{-3}$<br>(0.002) |
| Oversight orientation | | 0.006<br>(0.014) | 0.005<br>(0.005) | 0.001<br>(0.015) |
| Constant | 0.676***<br>(0.008) | 0.670***<br>(0.014) | 0.117***<br>(0.004) | 0.553***<br>(0.014) |

***Note.*** Ordinary least squares estimates; the equivalent ANOVA and ANCOVA statistics are reported in Online Appendix D. The DV in columns (1)–(2) is the error-detection rate (mean of the 10 per-task scores, each scored 0, 0.5, or 1); columns (3)–(4) re-estimate the column (2) specification with the false-positive rate and net detection (detection minus false positives) as DVs. The false-negative rate is one minus the detection rate, so its estimates mirror columns (1)–(2) with reversed sign. One observation per participant ($N = 400$). Self-explanation = 1 for the self-explanation condition and 0 for the system-provided-explanation condition. Standard errors in parentheses. * $p < .05$, ** $p < .01$, *** $p < .001$ (two-tailed).

*Supplementary Behavioral and Robustness Evidence*

Participants in the self-explanation condition recognized more of the three error categories than those in the system-provided-explanation condition (*Mean* = 2.04 versus 1.54). They were also more likely to identify all three categories correctly (28.5% versus 8.0%). False recognition of the three distractor categories that had not appeared was low and did not differ between conditions (*Mean* = 0.62 versus 0.71 of three), indicating that the higher recognition under self-explanation reflected more accurate discrimination (focal hits minus distractor false alarms: 1.42 versus 0.83 of three) rather than an indiscriminate tendency to endorse more categories. Pre-outcome recall elaboration was higher in the self-explanation condition than in the system-provided-explanation condition (*Mean* = 1.84 versus 1.43). Recall elaboration was positively associated with subsequent error detection among participants in the self-explanation condition (simple slope $b = 0.042$, $p < .001$), but not among those in the system-provided-explanation condition ($b = -0.002$, $p = .88$); the condition × recall interaction was significant ($p = .011$). Because recall was measured before the outcome tasks, this pattern provides convergent evidence that self-explanation produced a more accessible representation of verification-relevant reasoning before participants evaluated any new LLM output.

Consistent with this more accessible reasoning, self-explanation participants also spent more time reviewing the LLM-generated drafts, consulted reference materials more often, and made more extensive revisions than those in the system-provided-explanation condition (Table 4). Their false-positive rate was nonetheless lower (0.09 versus 0.12), indicating that this greater effort was discriminating rather than indiscriminate alteration of correct content. In the full-model specification, self-explanation reduced the false-positive rate by three percentage points (Table 5, column (3): $b = -0.030$, $p < .001$).

A net-detection measure (error detection minus the false-positive rate) yielded an even larger self-explanation advantage in the covariate-adjusted model (Table 5, column (4)). The result was also robust to dichotomizing error detection at the sample median: participants in the self-explanation condition remained far more likely to achieve high detection (odds ratio ≈ 3.9, $p < .001$), confirming that the effect did not depend on the graded scoring specification. Additional controls, alternative outcome and model specifications, and the equivalent ANOVA and ANCOVA statistics are reported in Online Appendix D.

Participants in the two conditions ultimately saw the same highlighted errors and the same standardized explanations; what differed was the work participants had done before the explanation appeared. The findings are therefore consistent with generative encoding making verification-relevant reasoning more accessible during later review. The supplementary behavioral results further indicate that the self-explanation advantage reflected more deliberate and discriminating verification rather than generalized distrust of LLM output. These behavioral indicators are theoretically compatible with the retrieval-based account but are not by themselves uniquely diagnostic of it: heightened attention and effort are the account's expected downstream expression rather than rival explanations, and the design-based mechanism evidence in Study 2 provides the sharper test.

## Study 2: Sustaining Error Detection Under Repeated Use

### *Participants and Design*

Study 2 was conducted with sales and pre-sales employees working in the same organization. These participants constituted a separate and non-overlapping sample from Study 1, and no participant took part in both studies, preventing learning spillovers across studies. Eligible participants met the same inclusion criteria as in Study 1 and were drawn from a candidate pool of 254 employees who had completed an internal firm survey and were active users of the firm's internal sales-support assistant during the study period. Participants were randomly assigned at the individual level to the retrieval-cue condition or the non-retrieval-cue condition. Assignment was performed administratively before the training session, and participants encountered their condition-specific procedures only at the end of that session, after the common baseline assessment (described below). Because the training session was scheduled as part of the department's regular work arrangements, participation was near-universal: 253 of the 254 eligible employees completed Study 2 (127 in the non-cue condition and 126 in the retrieval-cue condition). Thirteen participants who failed the same prespecified instructed-response attention checks were excluded prior to analysis (seven and six, respectively), resulting in a final analytic sample of 240 participants, with 120 participants in each condition. Participants in Study 2 were similar in age and tenure to those in Study 1. Completion was full: no participant in the analytic sample missed a study day or an embedded test email; the embedded emails arrived within participants' required daily workflow on scheduled workdays, so completing them did not depend on voluntary extra effort. Table 6 reports baseline descriptive statistics by condition.

**Table 6. Sample Descriptive Statistics for Study 2 at Baseline (Day 0), by Condition**

| Variable | Non-Cue | Retrieval Cue | Range | Variable | Non-Cue | Retrieval Cue | Range |
|---|---|---|---|---|---|---|---|
| Age (years) | 29.57 (6.35) | 30.97 (5.94) | 20.00–40.00 | Gender (F / M / NA) | 63 / 53 / 4 | 58 / 56 / 6 | — |
| Organizational tenure (years) | 4.88 (3.85) | 4.76 (2.82) | 0.00–15.00 | Error-detection rate (Day 0) | 0.62 (0.20) | 0.64 (0.19) | 0.00–1.00 |
| Objective AI knowledge (0/1) | 0.12 (0.33) | 0.18 (0.39) | 0.00–1.00 | False-positive rate (Day 0) | 0.10 (0.04) | 0.10 (0.04) | 0.03–0.27 |
| Domain capability (log sales) | 1.20 (3.50) | 0.57 (2.55) | 0.00–14.08 | Verification time (s, Day 0) | 132.20 (10.70) | 135.32 (9.78) | 100.40–157.60 |
| Oversight orientation (0/1) | 0.77 (0.42) | 0.70 (0.46) | 0.00–1.00 | | | | |

***Note.*** Cells report means with standard deviations in parentheses (counts for gender). The conditions were balanced on all baseline characteristics except verification time. Randomization tests: age $t(238) = 1.76$, tenure $t = -0.29$, objective AI knowledge $t = 1.25$, domain capability $t = -1.57$, oversight orientation $t = -1.17$ (all n.s.); gender $\chi^2(2, N = 240) = 0.69$, $p = .71$. The cue-condition difference in verification time at baseline is small (≈3 s; $t(238) = -2.38$, $p = .02$) and reflects a chance baseline imbalance rather than a treatment effect; baseline-adjusted analyses over Days 1–7 are reported in Online Appendix D.

### *Experimental Procedure*

Data collection for Study 2 occurred in two phases: an in-person training session followed by a seven-workday embedded task period (Figure 2b).

#### *Training Session and Baseline (Day 0)*

Because Study 2 examines how the stronger encoding mode identified in Study 1 can be sustained under repeated use, all participants attended the same training session on Day 0 and completed the same self-explanation exercise used in the self-explanation condition of Study 1 (described earlier); the design therefore does not include a parallel longitudinal system-provided-explanation arm, a boundary we revisit in the Limitations and Future Research section. Immediately afterward, all participants completed a common baseline assessment in which they reviewed and edited three LLM-generated draft responses using the same interface and reference materials. Because cue generation occurred only after this baseline assessment and cues were first displayed at login on Day 1, Day 0 provides a pre-treatment baseline measure of error detection.

*Embedded Simulated Task Period (Days 1–7)*

Following the training session, the study unfolded over the next seven workdays. Each workday, participants received their usual stream of customer inquiries, within which three simulated test emails were embedded per day and interleaved with normal messages to approximate routine work conditions. Participants were informed during consent that the study involved simulated customer inquiries embedded in their workflow during the study period, but they were not informed of the timing, frequency, or specific content of these test emails. The test emails (described earlier) were delivered through the organization's normal employee-facing workflow on a prespecified schedule; each contained two to four predefined errors. Backend research identifiers distinguished test emails from non-study messages in the log data. The test emails never reached customers or affected employee evaluation or service-level metrics.

Following the baseline assessment at the end of the training session, participants in the two conditions completed their respective customization activities in separate breakout rooms. In the retrieval-cue condition, participants generated a brief, self-chosen cue (a single word or short phrase) tied to the specific moment during training when they recognized how the LLM-generated response could be wrong. They were instructed not to restate a generic warning that an LLM can make mistakes or to formulate a general reminder to verify future outputs. Instead, they were asked to choose a word or brief phrase that they could associate with that realization episode and that could later reactivate the reasoning formed at that moment. The phrase could carry ordinary semantic meaning or be highly idiosyncratic (e.g., a personal keyword or even an arbitrary word) and did not need to be meaningful to others. Its self-generated association with the earlier reasoning episode defined it as a retrieval cue (see Online Appendix B for an illustration). Generated cues included, for example, “Matrix” and “pink elephant”; Online Appendix B reports a content analysis of all 120 cues. During the embedded task period, retrieval cues were delivered once per workday, beginning on Day 1, as part of the system’s standard login welcome message, which participants encountered at the start of the workday, not after each individual task. The cue was intentionally a lightweight, peripheral element of the workflow rather than a task-level prompt, minimizing disruption.

In the non-cue condition, participants were also asked to customize the same welcome message with a neutral or personally preferred word or phrase (e.g., a greeting or arbitrary term). However, this text was explicitly framed as a cosmetic customization, not a retrieval cue, and was not intended to trigger recall of the self-explanation episode. For example, some participants used “coffee ready” or “good morning.” All interface elements, timing of message presentation, and LLM drafting behavior were identical across conditions. The non-cue condition therefore functioned as a matched-message control: customization, visual presentation, daily exposure, and personalization were held constant. Assignment varied whether the personalized phrase was deliberately generated in connection with the earlier self-explanation episode or as a cosmetic customization. The phrases consequently varied in their surface semantics.

***Measures***

Error detection in Study 2 was scored from participants’ final outgoing responses to the embedded test emails (see the Measurements section): each predefined error was coded as corrected if the erroneous information was no longer present, and the email-level error-detection rate is the proportion of predefined errors corrected in that email. The next section describes how these email-level rates were aggregated for analysis and reports the alternative outcome codings used in the robustness checks, including a binary indicator of whether participants corrected the single most consequential error in each email, designated ex ante by domain experts as the predefined error posing the greatest compliance or financial risk to the customer. By design, exactly one error per email carried the highest such risk, so the designation involved no ties. The auxiliary verification measures defined in the Measurements section, together with a daily false-positive rate, were aggregated to the participant-day level.

***Data Analysis***

Each participant contributed 24 email-level observations across Day 0 and the seven embedded-use days, with three test emails per day, yielding 5,760 email-level observations. For the primary analyses, we computed each participant-day error-detection rate as the proportion of that day's predefined errors corrected across the three test emails, producing 1,920 participant-day observations from 240 participants.

We tested H2 using linear mixed-effects models with participant-level random intercepts. The models included study day, retrieval-cue condition, and their interaction as fixed effects, with objective AI knowledge, domain capability, and oversight orientation as covariates. Study day was coded from 0 to 7, and retrieval cue was coded 1 for the retrieval-cue condition and 0 for the non-cue condition. Because the two conditions received the same task set on each study day, day-specific task difficulty was shared across conditions and cannot account for the day × cue interaction; models incorporating day fixed effects provide an additional check. The day × cue interaction is the test of H2: a positive interaction indicates that error detection declined more slowly in the retrieval-cue condition than in the non-cue condition. We probed the interaction using condition-specific simple slopes. Supplementary analyses used alternative random-effects structures, email-level observations, models adjusted for Day 0 performance and estimated over Days 1–7, binary correction of the most consequential error, median-dichotomized error detection, false-positive rates, and net detection. Three of these alternative outcomes, namely the false-positive rate, net detection, and the share of most-consequential errors corrected, are reported alongside the primary models in columns (4) through (6) of Table 7; the remaining robustness checks appear in Online Appendix D. A sensitivity analysis based on the estimated precision of the interaction indicates that the design could detect a study day × retrieval-cue interaction as small as $b \approx 0.009$ per day with 80% power ($\alpha = .05$, two-tailed), comparable in magnitude to the observed coefficient.

Consistent with the causal logic of our two-study mechanism-testing strategy, we do not estimate an individual-level mediation model. The end-of-study cue-evocation measures are used as convergent process evidence and to evaluate competing explanations. Analyses were conducted in R 4.6.0 using maximum-likelihood estimation. Linear mixed-effects models used lme4 (v2.0-6), and fixed-effect tests used lmerTest (v3.2-1) with Satterthwaite degrees of freedom; ordinary least squares and logistic models used base R. All tests are two-tailed. Study 2 day-level correlations are reported in Online Appendix E, Table E2.

***Results***

*Randomization and Manipulation Checks*

Random assignment produced broadly comparable groups at the Day 0 baseline. The conditions did not differ in age, organizational tenure, objective AI knowledge, domain capability, oversight orientation, or gender, and Day 0 error detection was likewise balanced ($t(238) = 0.71$, $p = .48$). The retrieval-cue condition took about three seconds longer on verification at baseline ($t = -2.38$, $p = .02$); because this difference preceded cue generation and display, it reflects a chance baseline imbalance, and the day × cue interaction remains positive and similar in magnitude, although only marginally significant, in models that adjust for Day 0 performance (Online Appendix D). An exposure check confirmed that participants attended to the daily message: most participants in both conditions correctly recognized their own welcome phrase (86.7% in the retrieval-cue condition and 80.0% in the non-cue condition; the difference was not significant, $\chi^2(1, N = 240) = 1.92, p = .17$).

*Descriptive Trajectories of Error Detection and Verification Behavior*

We first examined model-free trajectories of error detection across study days. Figure 4a plots average error detection by condition with 95% confidence intervals. Error detection was broadly comparable between conditions at the Day 0 baseline (non-cue: Mean = 0.622; retrieval cue: Mean = 0.640; $t(238) = 0.71, p = .48$). Because Day 0 preceded cue generation and display, this small and nonsignificant gap is a random baseline difference. Because H2 concerns the difference in the rate of change across conditions, its test is the day × cue interaction rather than the Day 0 difference in levels. As the study progressed, however, clear divergence emerged across conditions. By the final embedded day (Day 7), mean error detection was 0.445 in the non-cue condition and 0.517 in the retrieval-cue condition. In the absence of a retrieval cue, error detection declined steadily over subsequent study days. Participants showed a gradual reduction in identifying erroneous statements in LLM-generated output, despite continued exposure to varied test emails and stable task structure.

A similar pattern was observed for verification behavior. As shown in Figure 4b, participants in the non-cue condition spent progressively less time reviewing LLM-generated responses and consulted reference materials less frequently across study days. These trends suggest that repeated LLM use under routine work conditions was accompanied by declining verification effort. We interpret this common decline as consistent with an erosion of oversight, although day-specific task composition cannot be completely separated from the shared temporal trend. The between-condition divergence, by contrast, is identified against each day's common task set and is therefore not attributable to task-specific difficulty (see also the day fixed-effects models in Online Appendix D). In contrast, when a self-generated retrieval cue was present, both error detection and verification behaviors were better sustained over time. Although some decline was observable, it was substantially smaller in the retrieval-cue condition. Because the day × cue interaction isolates the differential rate of change, the widening gap reflects the cue's effect on the trajectory of error detection rather than the small initial difference in performance.

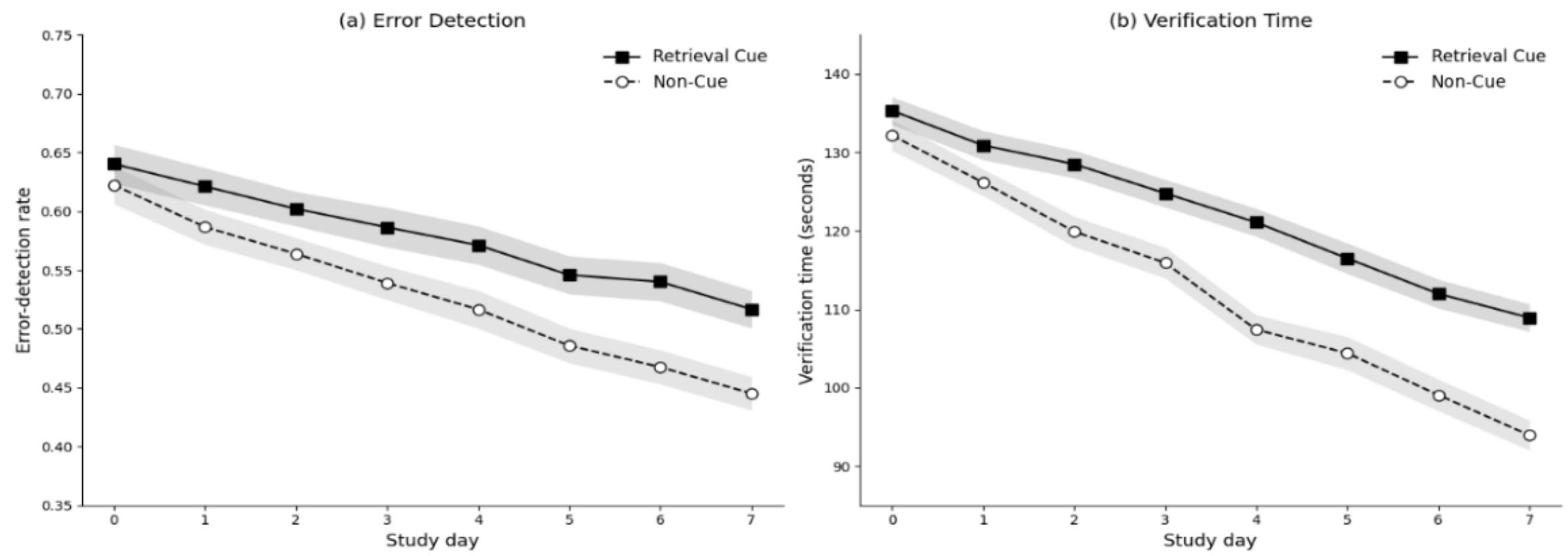

**Figure 4. (a) Error Detection and (b) Verification Time over Repeated LLM Use**

***Note***. Points are condition means at the participant-day level ($n$ = 120 per condition per day); shaded bands are 95% confidence intervals.

*Model-Based Test of H2*

We formally tested these descriptive patterns using linear mixed-effects models with random intercepts for participants. Error detection at the participant-day level served as the dependent variable, with study day, retrieval-cue condition, and their interaction included as fixed effects. Objective AI knowledge, domain capability, and oversight orientation were included as covariates. Error detection declined significantly over the study period. As shown in column (1) of Table 7, the fixed effect of study day was negative and statistically significant ($b = -0.021$, $p < .001$), indicating that error detection declined steadily across successive study days.

**Table 7. Mixed-Effects Regression Results for Study 2**

| | Error-detection rate | | | False positives | Net detection | Most consequential |
|---|---|---|---|---|---|---|
| Variable | (1) | (2) | (3) | (4) | (5) | (6) |
| Study Day | −0.021***<br>(0.002) | −0.025***<br>(0.002) | −0.025***<br>(0.002) | 0.005***<br>(0.001) | −0.030***<br>(0.002) | −0.101***<br>(0.004) |
| Retrieval Cue | | 0.023<br>(0.018) | 0.023<br>(0.018) | −0.001<br>(0.005) | 0.024<br>(0.019) | 0.042<br>(0.023) |
| Study Day × Retrieval Cue | | 0.008*<br>(0.003) | 0.008*<br>(0.003) | $0.070 \times 10^{-3}$<br>(0.001) | 0.008*<br>(0.003) | 0.021***<br>(0.005) |
| Objective AI Knowledge | | | −0.014<br>(0.020) | 0.003<br>(0.006) | −0.017<br>(0.021) | −0.023<br>(0.019) |
| Domain Capability (Log Sales) | | | −0.002<br>(0.002) | 0.001<br>(0.001) | −0.003<br>(0.002) | −0.003<br>(0.002) |
| Oversight Orientation | | | $-0.224 \times 10^{-3}$<br>(0.016) | 0.006<br>(0.005) | −0.006<br>(0.017) | −0.013<br>(0.016) |
| Intercept | 0.627***<br>(0.009) | 0.616***<br>(0.013) | 0.620***<br>(0.018) | 0.094***<br>(0.006) | 0.526***<br>(0.019) | 0.701***<br>(0.021) |
| Log-likelihood | 692.5 | 701.3 | 702.0 | 3248.8 | 572.8 | −238.5 |
| AIC | −1377.0 | −1390.7 | −1386.1 | −6479.7 | −1127.6 | 495.1 |

***Note.*** The unit of observation is the participant-day ($N$ = 1,920 observations from 240 participants), and models use maximum likelihood. All tests are two-tailed. All models include random intercepts at the participant level; the intraclass correlation from the unconditional means model is .27. The Retrieval Cue main effect is the model-implied contrast at Day 0 under the linear trend, not a balance test (see the Study 2 results section). Columns (1)–(3) use the daily error-detection rate as the DV; columns (4)–(6) re-estimate the full specification of column (3) with, respectively, the daily false-positive rate, net detection (detection minus false positives), and the share of most-consequential errors corrected. The false-negative rate is one minus the detection rate, so its estimates mirror columns (1)–(3) with reversed sign. Standard errors in parentheses; * $p < .05$, ** $p < .01$, *** $p < .001$.

Supporting Hypothesis 2, as shown in column (3) of Table 7, the interaction between study day and retrieval-cue condition was positive and statistically significant ($b$ = 0.008, p = .013). This interaction indicates that a self-generated retrieval cue mitigated the decline in error detection over time. Error detection decreased across study days in both conditions, but the rate of decline was substantially flatter for participants who received their cue. In simple-slope terms, error detection declined by $b = -0.025$ per day ($p < .001$) in the non-cue condition and by $b = -0.017$ per day ($p < .001$) in the retrieval-cue condition. In substantive terms, each day's three test emails contained nine embedded errors on average, so the cue's cumulative effect on the detection rate by Day 7 corresponds to catching about half an additional embedded error on the final day ($0.008 \times 7 \approx 0.056$). Equivalently, the cue offset about one third of the daily erosion (0.017 versus 0.025 per day).

Columns (2) and (3) of Table 7 also include a retrieval-cue main effect, which captures the difference between conditions at Day 0. Because Day 0 preceded cue generation and display, this term reflects a random baseline difference rather than an immediate cue effect. The model-based Day 0 contrast differs from the raw Day 0 comparison ($t(238) = 0.71$, $p = .48$) because the linear-trend specification uses information from all eight days. Consistent with this interpretation, when the model is re-estimated on Days 1–7 with Day 0 performance included as a baseline covariate, the day × cue interaction remains positive and similar in magnitude, although estimated less precisely and only marginally significant ($b = 0.007$, $p = .064$). The adjusted level difference is not significant ($b = 0.023$, $p = .23$; Online Appendix D). In the primary specification, the positive interaction shows that the cue significantly slowed the subsequent decline in error detection, so that the gap between conditions widened across repeated interactions instead of reflecting only a one-time boost in vigilance.

None of the covariates predicted error detection, and their inclusion did not substantively alter the estimated effects of study day or the retrieval-cue interaction. These results support Hypothesis 2. Columns (4) through (6) of Table 7 extend the column (3) specification to alternative outcomes. For the false-positive rate (column (4)), the day × cue interaction is essentially zero ($b = 0.070 \times 10^{-3}$, $p = .93$): false positives drifted up slightly over time in both conditions, and the cue neither amplified nor suppressed this drift. For net detection (column (5), detection minus false positives), the interaction remains positive and significant ($b = 0.008$, $p = .02$). For the expert-designated most consequential error in each email (column (6)), detection eroded most steeply ($b = -0.101$ per day) and the cue effect was largest ($b = 0.021$, $p < .001$). Together, these columns indicate that the cue selectively sustained the detection of genuine errors, including the errors that mattered most.

*Process Evidence for the Retrieval Mechanism*

Following the coordinated causal-chain strategy described earlier, we next examined whether the retrieval-cue intervention produced the proximal cognitive response specified by the proposed mechanism. This strategy calls for a coordinated pattern of evidence: the theoretically targeted intervention should affect the downstream behavioral outcome and should also affect a proximal indicator of the process through which that outcome is proposed to arise. The significant day × cue interaction reported above establishes the first part of this pattern: providing a cue associated with participants' earlier reasoning significantly slowed the decline in error detection over repeated use. We therefore examined whether the cue also evoked the verification-relevant reasoning with which it had been associated.

At the end of the study period, participants completed a brief probe asking what the daily welcome message brought to mind. Two coders blind to condition rated the responses on a 0–3 scale, where 0 indicated that the phrase brought nothing relevant to mind or was viewed as merely decorative, and 3 indicated an explicit reminder that LLM output could contain errors and should be verified. Intercoder agreement was high (Cohen's $\kappa = .83$; linear-weighted $\kappa = .88$). Evocation ratings were substantially higher in the retrieval-cue condition than in the non-cue condition (*Mean* = 2.15, *SD* = 0.83 versus *Mean* = 1.36, *SD* = 0.96; $t(238) = 6.85$, $p < .001$). Similarly, 77.5% of participants in the retrieval-cue condition produced responses reflecting a verification mindset (coded evocation rating $\geq 2$), compared with 48.3% in the non-cue condition.

A separate cue–episode association scale provided further evidence that the cue reinstated the earlier encoding episode. The association composite was higher in the retrieval-cue condition than in the non-cue condition (*Mean* = 2.86 versus 2.44; $t(238) = 6.98$, $p < .001$; Cronbach's $\alpha$

= .81). The same pattern emerged for the item asking whether the daily message brought the earlier self-explanation episode to mind. Conversely, participants in the non-cue condition were more likely to interpret their phrase as merely decorative (non-cue M*ean* = 3.74 versus retrieval cue *Mean* = 3.51 on the decorative item, coded so that higher values indicate a more decorative interpretation). Full item wordings appear in Online Appendix C. Because the non-cue condition received an equally personalized message through the same channel, the experimental contrast tests the effect of deliberately generating the personalized phrase in connection with participants' prior verification-relevant reasoning rather than as a cosmetic customization. Additional tests of message attributes are reported in the Robustness Checks section below.

As supplementary consistency evidence, coded evocation ratings were associated with flatter error-detection trajectories across the full sample (day × evocation: $b = 0.006$, $p < .001$). Participants whose daily message more strongly evoked verification-relevant reasoning exhibited less deterioration in error detection over time, mirroring the randomized experimental contrast. We do not interpret this association as an individual-level mediation test. The evocation measure was collected only at the end of the study, after the repeated behavioral outcomes, and offered limited within-condition variation. It is therefore treated as a manipulation-linked process indicator rather than as a temporally situated mediator or a direct measure of retrievability.

The causal effect of the retrieval cue on the error-detection trajectory is identified by the randomized experimental contrast, whereas the mechanism inference rests on the coordinated pattern of evidence specified by our two-study mechanism-testing strategy. The theoretically targeted retrieval-support intervention both evoked the prior reasoning with which the cue had been associated and produced the predicted change in the subsequent behavioral trajectory. Taken together with the matched-message design, this pattern is consistent with the cue operating by reactivating prior verification-relevant reasoning. The cue did not repeat the original explanation or provide task-specific corrective information.

More broadly, these results complement the mechanism evidence from Study 1. Study 1 showed that generative encoding improved pre-outcome recall and subsequent error detection, indicating that oversight-relevant information was initially encoded in a more retrievable form. Study 2 held generative encoding constant and provided evidence that a cue associated with the earlier reasoning can help sustain access to that reasoning and slow the erosion of error detection during repeated use. The two studies therefore provide coordinated evidence for the two theoretically specified components of the retrieval process: establishing retrievability through initial encoding and sustaining access through subsequent retrieval support.

*Robustness Checks*

The day × cue interaction was positive and similar in magnitude across alternative specifications and outcome operationalizations. It remained statistically significant in random-slope, day-fixed-effects, email-level, error-level, and net-detection models. When Day 0 performance was included as a baseline covariate in models estimated over Days 1–7, the interaction remained positive but was estimated less precisely. Binary correction of the single most consequential error yielded the strongest form of the pattern: detection of the expert-designated highest-risk error in each email declined most steeply and showed the largest cue effect. Median-dichotomized error detection yielded the same substantive conclusion (Online Appendix D).

The results offer little support for the possibility that the retrieval cue merely induced generalized skepticism toward LLM output. A net-detection measure, calculated as error detection minus false positives, reproduced the focal pattern. False-positive rates were nearly identical across conditions (0.117 versus 0.116), and models predicting false-positive outcomes showed no corresponding cue-induced increase. Together, these results are more consistent with sustained detection of genuine errors than with indiscriminate challenge of LLM output.

The cue texts also provide evidence against a content-only account of the trajectory effect. Excluding the three participants whose cues departed from the generation instruction by stating a general fallibility proposition, the day × retrieval-cue interaction remained positive and significant. Within the 117 instruction-compliant cue users, the idiosyncratic subgroup showed a flatter estimated trajectory than the non-cue condition, and its slope did not differ detectably from that of cues carrying ordinary semantic meaning. The treatment effect survives removal of the noncompliant generic statements. Cues carrying ordinary semantic meaning show no detectable slope advantage over idiosyncratic cues, a pattern that weighs against a strong content-only explanation (Online Appendix D, Figure D1).

Finally, placebo checks offered little support for an account based on measured generic attributes of the daily message. The non-cue condition received an equally personalized message in the same visual format and through the same channel, and participants in both conditions showed comparable recognition of their own welcome phrase. The conditions also did not differ in perceived personal meaningfulness, visual noticeability, memorability, liking, or attention capture; the composite of these message attributes did not differ significantly between conditions ($t(238) = 1.34$, $p = .18$). Moreover, these attributes did not predict error-detection trajectories, either across the full sample or within the retrieval-cue condition. Together, these checks offer little support for explaining the trajectory difference simply in terms of participants receiving a more noticeable, memorable, or appealing personalized message.

## Overall Discussion

Both studies support the same retrieval-based account: oversight can fail when relevant information is not accessible at the moment of review, even in settings where reviewers have received that information and are actively reviewing the output. The results indicate that generating an explanation supported later accessibility of verification-relevant reasoning and that an associated cue helped reactivate that reasoning across repeated use. Building on these findings, we develop three contributions in dialogue with capability- and engagement-based accounts of human oversight.

***Retrievability as a Distinct and Complementary Precondition for Oversight***

Our first contribution is to identify retrievability as a distinct precondition for effective oversight. Existing accounts attribute oversight failure either to limited literacy, expertise, or mental models, motivating interventions that build these resources [2,20,39], or to insufficient scrutiny of seemingly reliable systems, motivating interventions that increase attention or evaluative effort [6,28]. Neither guarantees that the relevant information is accessible when a particular output is reviewed; our results identify retrievability as an additional step at which oversight can fail. This condition changes how capability and engagement relate to oversight. Capability concerns what users know and can do, whereas engagement concerns whether they undertake and sustain review. Neither ensures that the reasoning needed for a particular review will come to mind. A user may therefore be capable and engaged yet still fail to detect an error because the relevant reasoning remains inaccessible at that moment. Retrievability is not simply a third input into oversight; it is a condition under which capability and engagement can be translated into effective oversight.

This argument connects human oversight to adjacent research. Judgment research shows that the influence of information depends in part on its accessibility at the time of judgment [12], whereas research on professional verification shows that the accessibility and organization of expert knowledge shape the quality of checking [14]. We bring this insight into human oversight by treating retrievability as an independent precondition rather than assuming that relevant information will be made available by capability and engagement. LLM-generated text makes this accessibility problem especially consequential. Unlike a bounded recommendation or score, a fluent draft can embed many claims in a coherent whole, remain mostly correct, and provide little indication of which claim warrants scrutiny or what knowledge should be brought to bear on it. Under high-volume, routine use, effective review depends on retrieving the right reasoning for the right claim at the right moment [7,28,37]. Even when broad classes of LLM errors are predictable, which claim will be erroneous in a particular draft may not be, and eliminating all consequential errors ex ante may be infeasible or prohibitively costly. Although our studies operationalize oversight for LLM-generated text, the retrievability precondition itself is not specific to LLMs: any setting in which users review automated decision input—recommendations, risk scores, or generated content—raises the same question of whether previously encoded verification-relevant reasoning becomes accessible at the moment of review. The theoretical implication is that oversight support must address not only what users know and whether they review, but also whether relevant reasoning is accessible in the specific review context.

***Encoding Oversight-Relevant Information for Later Retrieval***

Our second contribution is to explain why informationally equivalent oversight support can produce different levels of error detection. Research on explanations has located their effectiveness in the information they provide, asking how they can be made more complete, accurate, or useful [3,13,16]. This perspective explains why explanatory content matters, but not why users who receive the same information may later differ in their ability to apply it. Our results show that what users do with information as they learn it can matter independently of what the information contains. Encoding is thus a separate target of oversight support. Content determines what information is available to users, whereas encoding shapes the representation that remains available for later review. Oversight support may therefore be informationally sufficient yet functionally weak if it leaves users with information that is difficult to access when they confront a new output. Conversely, later oversight may improve without adding information when users engage in processing that prepares reasoning for later retrieval.

This argument connects human oversight to the generative-learning tradition while extending its application. Generative-learning research shows that producing an explanation, rather than only receiving one, can improve comprehension and retention [4,9,10]. Oversight introduces a further challenge: users must apply reasoning formed in one encounter to detect errors in new outputs. Our account proposes that generation supports this transfer because constructing an explanation engages operations that resemble those required during later review. The resulting reasoning is therefore more likely to be reinstated by cues in the review context [27,36]. Self-generated explanation is one instance of a broader encoding-side design principle: engage users in representational work that prepares verification-relevant reasoning for later retrieval. Other techniques that elicit comparable processing may serve the same function. Taken together, the theoretical shift is from assessing oversight support only by the information it delivers to also considering the representation it leaves behind. What users are told matters, but so does how that information is encoded for use in a later review.

***Sustaining Oversight by Reactivating Prior Reasoning***

Our third contribution is to develop and provide evidence for reactivation as a mechanism for sustaining oversight as error detection declines across repeated use. Interventions delivered at the point of use commonly add something to each encounter, such as an explanation, performance information, a reflection prompt, or a cognitive forcing function [1,3,6,42]. These interventions can support an individual review, but they provide a limited account of what happens as users repeatedly interact with the same system and deliberate checking gives way to routine. We thus extend theory beyond the isolated or short-horizon context of most of the available oversight research.

A retrieval-based account offers a different explanation for this decline. Reasoning that was accessible when first formed may become less likely to come to mind as repeated interactions become familiar. Oversight may weaken even when users retain the relevant information and remain willing to review the output. Reactivation addresses this problem by restoring access to reasoning users have already formed. An associated cue can sustain error detection without retraining, repeating the original explanation, or supplying task-specific corrective information. This account distinguishes reactivation from a generic increase in attention. A cue matters not simply because it is noticeable, personalized, or cautionary, but because it has become linked to prior verification-relevant reasoning. Its function is to bring that reasoning back into the review context. This also distinguishes reactivation from blanket skepticism: restoring access to diagnostic reasoning should preserve users' ability to distinguish erroneous from acceptable outputs rather than merely encouraging them to reject more outputs. This argument links oversight to theories of retrieval and forgetting, in which access to acquired knowledge can decline with disuse but be restored by cues aligned with its original encoding [36,40]. Whereas this literature has largely examined memory for facts and skills, we bring its temporal logic into human oversight. Declining detection under routine use can be understood partly as a retrieval problem and addressed through reactivation rather than re-instruction.

The theoretical shift is from treating oversight as something established at a single point in time to treating it as something that must be maintained across repeated encounters. Oversight support should address not only the quality of an individual review, but also whether users' prior verification-relevant reasoning remains accessible as interaction with the system becomes routine.

### *Practical Implications*

Our findings show a way of supporting human oversight of LLMs in routine work. Improving oversight can stem from changing how the information reviewers already have is encoded and later brought back to mind, without repeating the earlier explanation or adding task-specific corrective information at each encounter. For organizations, this reframes oversight support as design at two moments. At onboarding, have users explain why an example output is wrong, and how that would be verified. Generating their own reasoning, rather than only receiving the answer, can make it more retrievable later. In the workflow, capture a brief cue the user associates with that reasoning and surface it where later outputs are reviewed, so the cue can help the user reinstate the reasoning without needing to be retaught. Because gains erode as review becomes routine, treat oversight as something to sustain, tracking error detection alongside false positives and review time rather than assuming a single training session resolves it. One caution follows from the account: retrieval support lowers the informational barrier to oversight but does not remove its cost. Careful review consumes attention and time, and adding oversight duties, without adjusting output targets, risks trading caught errors for pace [7]; retrieval-oriented support belongs alongside realistic expectations, not layered atop unchanged ones. Individual reviewers can apply the same logic by explaining an error in their own terms and keeping a self-chosen cue. The cue can support later detection when the original explanation is no longer present. For those who build LLM tools, it suggests designing for retrieval rather than only for disclosure: prompting users to generate their own account of an error, and re-presenting user-associated cues at review, instead of relying only on warnings and disclaimers.

***Limitations and Future Research***

Our design leaves several questions open, each a natural next step. First, our experiments manipulate antecedents posited to support retrieval and document their behavioral consequences, but they do not observe retrieval as it unfolds. Future work could track whether the relevant reasoning is accessible immediately before a review, which would add temporal resolution to the account. Second, reactivation may operate partly through the attention and effort it recruits. Our matched-message control and message-attribute checks address personalization and measured salience, memorability, and liking, but they do not fully separate retrieval from momentary attention or effort; quantifying these downstream channels is a question for future work. Third, we studied one organization, with searchable reference materials and a fixed set of error types; how much retrieval support helps may differ for tasks that lean on deeper expertise, for higher-stakes decisions, and for settings where no reference material is at hand. Fourth, future research should examine cue frequency and specificity, whether model updates change their relevance, and what restores their effectiveness as it weakens over time. Our design also isolates one link of the process at a time in a real setting; a complementary approach we did not take would manipulate capability, engagement, and retrieval support together to map how the three preconditions combine. Such manipulation might need to take place in a laboratory setting, in which a repeated-usage design might not be possible. Finally, because every participant in Study 2 began from the same self-generated explanation, we could follow the durability of that stronger mode cleanly but could not observe a provided-explanation group over the same period; a longer study that follows both would show how the initial gap in encoding evolves with continued use.

## Conclusion

Effective review depends in part on whether oversight-relevant information is accessible at the moment of review. Our field studies support a retrieval-based account of oversight failure: generating one's own explanation strengthened the encoding of verification-relevant reasoning, and as LLM use became routine, a self-chosen cue helped keep that reasoning accessible. Effective LLM oversight cannot be secured once at training; the reasoning that verification requires must remain recoverable when it is needed. Designing workflows for retrieval can therefore complement investments in user capability and engagement.

# Appendix

## Appendix A. Extended Review of Related Literature on Human Oversight of AI Output

This appendix provides the extended review that the literature review of the main text summarizes. The two sections below present the capability-based and engagement-based explanations in detail; the table below then organizes representative studies by the mechanism they examine.

### *The User-Capability Explanation in Detail*

The first explanation locates oversight failure in capability, attributing it to deficiencies in the accurate and usable evaluative resources a user can bring to an AI output [15,18,38]. These resources concern either the AI system or the substantive task. On this account, oversight fails when users lack the relevant resources, hold inaccurate or incomplete representations of the system or task, or lack the procedures needed to distinguish correct from erroneous content. The explanation therefore speaks directly to error detection. A user who cannot anticipate where a system errs, cannot recognize a false claim, or cannot verify a suspicious one will pass errors through review even when they examine the output.

The first category, AI-specific resources, concerns whether users understand what a system can and cannot do and how much evidentiary weight its output warrants. AI literacy provides a general competence to understand, use, and critically evaluate AI across systems [38], whereas mental models represent more system-specific beliefs about how a particular system performs, the limits of its competence, and how it tends to fail [15]. Limited AI-specific capability produces recognizable failures of oversight. Users may not appreciate that high average accuracy does not imply reliability on any given case, may hold miscalibrated beliefs about those limits [43], or may treat fluent and coherent output as evidence of correctness. In each case they misjudge which claims warrant scrutiny and what errors to anticipate, so plausible errors are never flagged for checking.

Interventions aimed at strengthening AI-specific capability seek to supply, clarify, or correct users' understanding of the system and its use. They include AI-literacy training and disclosures of system limitations, performance and uncertainty information intended to calibrate reliance, and system-provided explanations that make the basis or sensitivity of an output more interpretable [13,16]. These interventions can work, but not uniformly. Explanations that reveal genuine discrepancies between human and system reasoning can improve users' metacognitive accuracy by helping them assess their own ability relative to the AI, reducing overconfidence and improving delegation decisions [43]. Yet feature-based explanations also reshape how users weight evidence and may entrench rather than correct an inaccurate mental model [3], and performance information improves calibrated use only when it is presented and processed in ways users can interpret appropriately [42]. Interventions aimed primarily at increasing attention or evaluative effort, such as generic warnings and cognitive forcing functions, are more directly tied to the engagement-based account below, though some may operate through both pathways when a warning also conveys new information about a system's limits or error patterns.

The second category, task-specific resources, concerns whether users command the domain expertise, reference evidence, and verification procedures needed to verify whether the content of a particular output is correct. Even a user who understands a system's fallibility cannot catch a substantive error without the means to adjudicate the claim itself. Domain experience does not act through a single pathway. Task-based experience strengthens users' ability to review and complement AI output and to recognize inaccurate predictions [2,39], and human intervention improves machine predictions only when the human contributes genuinely useful complementary judgment [30]. Broader professional seniority, by contrast, can shape willingness to rely on the system rather than the ability to review it, a distinct engagement pathway considered below. The task side carries its own limits. The reference evidence against which output must be verified may itself be incomplete, ambiguous, or contested [19].

GenAI reframes how these two forms of capability must be combined. A conventional decision aid presents a bounded recommendation, score, or diagnosis that the user accepts or rejects. Generated content instead consists of multiple claims embedded in a coherent whole, and a draft may be globally plausible and mostly correct yet contain one consequential local error [7]. Effective review therefore requires the user to decompose the output, identify the claims that warrant verification, locate relevant evidence, and correct the erroneous content without discarding the valid. This demands AI-specific understanding of why fluent, confident output can still be wrong together with the task-specific means to test a particular claim. How users should be equipped to review generated output, and how those resources should be built, remains far less understood than in the decision-making setting. As defined in the main text, oversight-relevant information is the content within these resources that bears on reviewing a given output—in brief, whatever a reviewer would need to have in mind in order to recognize that a particular claim in a particular output is wrong. Capability-based accounts explain whether users possess adequate oversight-relevant information. They do not explain whether users undertake and sustain the evaluative activity needed to apply it, which is the question the engagement-based account takes up next.

### *The Engagement Explanation in Detail*

The second explanation locates oversight failure in engagement, attributing it to failures to initiate, appropriately allocate, or sustain the attention and evaluative effort needed to scrutinize AI output and regulate reliance on the system [28,32]. Users may hold the information and procedures needed to identify an error yet fail to examine the relevant claim independently, compare it against evidence, or sustain monitoring of a system that usually performs well. Errors then survive not because users lack the ability to detect them, but because the evaluative activity required for detection is not undertaken or sustained.

A first engagement problem concerns whether users initiate and appropriately allocate independent scrutiny rather than defaulting to the system's output. A large literature on automation bias documents that users substitute automated recommendations for their own information processing, producing both omission errors, when they miss a problem the system did not flag, and commission errors, when they follow an incorrect recommendation [32]. The same tendency appears with AI recommendations, which can encourage fast, intuitive processing in place of deliberate review [25], and users sometimes follow clearly erroneous AI advice even when they are capable of reaching the correct decision independently [22]. In each case the mechanism is not that users lack the relevant resources but that AI output becomes the default response, displacing the independent review they could have performed.

A second engagement problem concerns whether users sustain scrutiny and regulate reliance over repeated interaction. When a system usually performs well and errors are rare, the immediate return to continued monitoring is low, and users gradually reduce the effort they invest, a complacency that leaves the occasional error unexamined [28]. Effective oversight also requires reliance calibrated to the system's actual reliability [17,21]. Reliance can diverge from appropriate levels in both directions [17], but overreliance is more central to the present problem of detecting errors in AI output. Users who place excessive weight on AI output subject it to less independent verification, allowing erroneous content to pass through review. Underreliance creates a different problem by causing users to discount or reject accurate advice [17].

Interventions aimed at strengthening engagement seek to interrupt default acceptance and raise the likelihood that users examine AI output independently. Cognitive forcing functions require users to form or explain their own assessment before seeing or accepting the system's recommendation, and they reduce overreliance relative to standard displays, though users find the more effective versions less agreeable [6]. Design features that prompt reflection during the task can similarly induce users to reconsider an AI-supported decision [1]. Because these interventions primarily target attention and effort, they belong to the engagement-based account; confidence and uncertainty displays sit closer to the boundary because they also convey information about system reliability. Performance information likewise improves calibrated use only under particular forms of presentation and processing [42]. These interventions can increase independent scrutiny, but they impose effort and may be resisted or inconsistently followed. Moreover, greater overall scrutiny does not by itself guarantee that users will apply the particular information or reasoning needed in a given review.

GenAI systems intensify both engagement problems. A generated draft is fluent, extended, and delivered in a confident register. When such output is mostly correct and errors are infrequent, exhaustive verification is costly and the immediate payoff from checking any given claim is low [7]. As repeated use makes review increasingly routine, users have strong incentives to accept plausible content, selectively reduce scrutiny, and treat verification as a formality [28,37]. In this setting, the engagement challenge is whether users initiate claim-level scrutiny and sustain sufficient evaluative effort across repeated use, particularly when only a small portion of the output warrants concern.

Engagement-based accounts explain whether users initiate and sustain the evaluative activity needed to apply oversight-relevant information. Together with capability-based accounts, they describe whether users possess the relevant evaluative resources and whether they actively review the output. These accounts do not, however, explain whether the particular oversight-relevant information required for a particular output becomes accessible during that review. A user may know that the system can err and may be actively scrutinizing its output, yet still fail to retrieve the particular domain knowledge, error pattern, or verification strategy needed to recognize the current error. This residual possibility, present even when users possess the relevant resources and actively review the output, is the gap the main text develops.

| Appendix Table A1. Related Literature on Human Oversight of AI Output | | | |
|---|---|---|---|
| **Specific focus or mechanism** | **Study** | **Context and method** | **Main findings** |
| ***Capability Perspective*** | | | |
| AI literacy | [38] | Scale development and validation; multiple samples | Develops and validates a multidimensional measure of users' knowledge and skills for understanding, using, detecting, and evaluating AI. |
| Explanation-supported understanding | [16] | Theoretical synthesis of explanations from intelligent and knowledge-based systems | Well-designed, context-specific explanations can support understanding, performance, trust, and acceptance, although their effectiveness depends partly on the cognitive effort required to use them. |
| Domain experience and complementary ability | [2] | Within-subject field experiment; corporate IT-support workers resolving service problems with and without an algorithmic tool | Domain experience increases workers' ability to complement algorithmic advice, including identifying inaccurate predictions, but very high experience can also increase aversion to accurate advice; workers with moderate experience benefit most from the algorithmic tool. |
| Task-based experience | [39] | Field study; knowledge workers using an AI system for medical chart coding | AI produces greater productivity gains for workers with more task-based experience, indicating that task-specific knowledge helps workers benefit from AI; senior workers gain less, partly because of lower trust in the system. |
| Domain expertise and system understanding | [18] | Physicians using an AI diagnostic system; experimental/process evidence | Inadequate monitoring of one's own reasoning and the AI system can lead professionals to accept incorrect diagnostic advice. |
| Complementary human expertise | [30] | Field experiment; retail demand forecasting | Human intervention improves machine-learning predictions only under conditions in which humans possess useful complementary judgment. |
| Metaknowledge for delegation | [15] | Multi-study experiments; human–AI task delegation | Humans hold inaccurate metaknowledge about their own and the AI's capabilities, which limits productive delegation and the value they add in collaboration. |
| ***Engagement Perspective*** | | | |
| Automation bias | [32] | Laboratory experiments with automated decision aids | Users commit omission and commission errors when they substitute automated recommendations for independent information processing. |
| Automation complacency and attention | [28] | Integrative theoretical review | Explains automation bias and complacency as failures to allocate and sustain attention during monitoring and decision making. |
| Trust calibration | [21] | Integrative theoretical review | Effective human–automation interaction requires reliance calibrated to the system's actual capabilities and reliability. |
| Algorithm | [42] | Multi-study experiments | Providing performance information does not uniformly improve calibrated |
| appreciation | | | use; its effect depends on how the information is presented and processed. |
| Algorithmic conformity | [22] | Four experiments | Users follow clearly erroneous AI advice despite being capable of answering correctly, partly because of discomfort with disagreeing with AI. |
| Fast cognition | [25] | Multi-study experiments | AI recommendations encourage fast processing and increase the sharing of both true and false information. |
| Cognitive forcing | [6] | Experiments with AI-assisted decision making | Cognitive forcing functions reduced overreliance relative to standard explanations, although users rated the more effective interventions less favorably. |
| Machine-induced reflection | [1] | AI-augmented diagnostic work; experiments | AI design can trigger reflection on a current decision, but the study does not examine retrieval of previously encoded oversight-relevant information. |
| Aversion and appreciation | [17] | Integrative theoretical review | Integrates algorithm aversion and appreciation as patterns of under- and over-reliance, framing calibrated engagement with algorithmic advice as the central challenge. |
| Evaluative know-how under opacity | [20] | Field study; medical diagnosis professionals using AI tools | Professionals confronting opaque AI develop interrogation practices that draw on their own domain know-how; when those resources fall short, they disengage from the AI's input. |
| ***Retrievability Perspective*** | | | |
| Accessibility of previously encoded oversight-relevant information | Current study | Two lab-in-the-field experiments involving organizational users reviewing LLM-generated output | Users may possess the information needed to detect an LLM error and actively review the output, yet fail to apply that information when it is not accessible at the point of review. Self-generated explanation improves later unaided error detection, and a self-generated retrieval cue slows the decline in detection over repeated use without repeating the original explanation or supplying task-specific corrective information. |

***Note.*** Effective oversight may also depend on artifact and organizational conditions, such as whether a system exposes diagnostic evidence and whether users have authority to intervene. This table focuses on the user-side preconditions that determine whether an error is detected at review. Studies are grouped by the mechanism most relevant to the three preconditions. Some studies also identify mechanisms tied to another precondition. For example, several capability studies report findings that also involve trust calibration or advice aversion.

## Appendix B. Illustration of the Self-Generated Retrieval Cue Mechanism

To clarify how the retrieval-cue intervention functioned in practice, we provide an illustrative example based on typical participant responses. During the initial self-explanation exercise, participants reviewed LLM-generated draft responses to customer inquiries with reference materials and were asked to articulate, in their own words, what might be wrong with the response and how such an error could arise. For example, a participant might observe that the system appeared to apply configuration rules from an outdated product version and write:

> *"The AI probably generalized from older documentation and ignored version constraints. I should always check applicability conditions instead of trusting fluent wording."*

Immediately afterward, participants assigned to the retrieval-cue condition were asked to generate a brief word or phrase that would later help them recall the moment they realized the LLM could be incorrect. Such a participant might then select the cue "version trap."

During the subsequent seven-day embedded task period, this cue was displayed once per workday as part of the system's standard login welcome message (e.g., "Welcome back — reminder: version trap."). The cue did not repeat the original explanation or provide task-specific corrective information. It could carry ordinary semantic meaning, but its defining feature was its self-generated association with the participant's earlier realization episode. It was therefore designed to reinstate that episode and make the logic the participant had previously articulated more accessible when reviewing new drafts.

Participants generated brief cues such as "yellow light," "version trap," and "familiarity trap," as well as highly idiosyncratic labels such as "pink elephant" and "2:34" that were meaningful only to their authors but became associated, through self-generation, with the cognitive state of discovering an error during training. Every retrieval cue was presented in the same peripheral manner and was not linked to any particular task, document, or error category. The instruction anchored each self-generated phrase to the participant's own realization episode rather than asking the participant to restate the generic proposition that an LLM can make mistakes.

To characterize what participants generated, we classified the 120 retrieval cues by form and strict compliance with the generation instruction (Table B1). Cues were short (Mean = 4.7 characters, SD = 2.0, range 1–15); 9.2% used an explicit negation or contrast marker (≠, "don't"), and 8.3% were phrased as questions. Fifty-two cues (43.3%) were idiosyncratic personal tags without a direct verification instruction, and 65 (54.2%) were brief cues with ordinary meaning selected at the realization episode. Only three cues (2.5%) restated a general proposition about AI fallibility rather than using an episode-linked label and were coded as strict departures from the instruction. Thus, 117 of 120 cues (97.5%) followed the cue-generation instruction.

By contrast, the messages customized in the non-cue condition were greetings, status lines, or motivational phrases (e.g., "good morning," "state: ON," "coffee ready," "progress bar +1"), and none contained verification-related content. This contrast is consistent with the matched-message design. Both conditions received equally personalized daily messages, and only the retrieval-cue condition's messages had been associated with prior verification-relevant reasoning. Table B1 describes cue form and strict instruction compliance, not mutually exclusive psychological mechanisms. Ordinary meaning and episode-specific association can coexist. Appendix D reports an exploratory comparison of trajectories across the two instruction-compliant cue forms.

**Appendix Table B1. Form and Instruction Compliance of the 120 Self-Generated Retrieval Cues**

| Category | Description | Illustrative cues (English translation; Chinese original) | n (%) |
|---|---|---|---|
| Idiosyncratic personal tags | Private images, metaphors, timestamps, or personal labels meaningful mainly to their author | "2:34"; "pink elephant" (粉色大象); "two moons" (两个月亮) | 52 (43.3) |
| Instruction- | Brief labels with ordinary | "yellow light" (黄灯); "version trap" (版 | 65 (54.2) |
| compliant cues with ordinary meaning | meaning selected for association with the participant's realization episode | 本陷阱); "familiarity trap" (熟悉感陷阱) | |
| General fallibility statements | General propositions rather than episode-linked labels; strict departures from the cue instruction | "professional tone ≠ evidence" (专业腔≠证据); "AI is not all-powerful" (AI 不是万能的); "confidence ≠ correctness" (自信≠正确) | 3 (2.5) |

Note. The first two rows comprise the 117 cues (97.5%) that followed the instruction to select a brief label associated with the participant's realization episode; the final row contains the only three cues (2.5%) coded as strict departures. Categories describe cue form rather than mutually exclusive psychological mechanisms. English translations are illustrative; participants wrote cues in Chinese, occasionally with English words or symbols. Percentages sum to 100.

## Appendix C. Measures and Materials

Behavioral outcomes (error detection, false-positive rate, verification time, reference-material consultation, textual modification) were computed from system logs and are defined in the measurement sections of the main text. Reliabilities are from the analytic samples (Study 1 $N$ = 400; Study 2 $N$ = 240). Please note that all materials were administered in Chinese; English is an illustrative translation.

### *Study 1 measures*

**Manipulation check (encoding source).** Single forced-choice item: "How were the explanations for the examples produced?" (generated my own first, then saw feedback / provided to me directly); 「示例的解释是如何产生的？」 94.5% and 91.0% correct across conditions. Developed for this study.

**Comprehension check.** Three error-category items (count 0–3): "Which categories of errors appeared in the examples?"; 「示例中出现了哪些类型的错误？」 Six options were presented: the three focal categories (applicability or condition error 「适用性或条件错误」, constraint or exception error 「约束或例外错误」, procedural error 「流程错误」) and three distractors (numerical or calculation error 「数字或计算错误」, source or citation error 「来源或引用错误」, tone or wording error 「语气或表达错误」); participants could select any number. Developed for this study.

**Recall elaboration.** Open-ended response to the prompt "In your own words, please describe the errors you saw in the examples and what you took away from them"; 「请用你自己的话描述：你在刚才的示例中看到了哪些错误？你从中获得了什么收获？」 Coded 0–3 (0 = no error mentioned … 3 = mechanism/lesson articulated); inter-rater $\varkappa$ = .82, linear-weighted $\varkappa$ = .86.

**Objective AI knowledge (covariate).** Single binary item: whether the respondent correctly identified the primary application domain of AI (1 = correct, 0 = otherwise); item wording appears under the objective AI-knowledge test below. Administered in both studies.

**Domain capability (covariate).** Computed from administrative sales records rather than a survey item: log(1 + sales-deal amount) for the focal product line, with participants recording no sales during the one-month pre-experiment window assigned a value of zero before the transformation. Available for both studies.

**Oversight orientation (covariate).** Binary indicator from a multi-select item on reasons for not accepting AI-assisted selling: coded 1 if the respondent selected "the accuracy and correctness of AI work cannot be guaranteed for the company's product system" 「公司的产品体系辅助，无法保证 AI 的工作是准确和正确的」 or "the quality, validity, and reliability of AI models are unclear" 「不清楚人工智能模型的质量、有效性和可靠性」, and 0 otherwise. The full item read "What are the main reasons you would not accept an AI tool to assist your sales work? (select all that apply)"; 「您不接受人工智能（AI）工具辅助您完成销售工作，主要的原因是什么？（多选）」 with nine options: limited practical help in advancing projects 「AI 作用有限，无法帮我实质性推动项目进程」; unclear model quality, validity, and reliability 「不清楚人工智能模型的质量、有效性和可靠性」 (scored); unsatisfactory past answers and not knowing how to prompt 「不知道如何提问，以前用过，但是得到的答案不太满意」; no organizational arrangement for effective use 「公司并没有统一安排，不知道该如何有效使用」; accuracy cannot be guaranteed for the company's complex product system 「公司的产品体系辅助，无法保证 AI 的工作是准确和正确的」 (scored); not knowing which solution to evaluate or select 「我们不知道要评估/选择哪个解决方案作为 AI 应用」; concern about information leakage, especially deal information 「担心使用工具导致信息泄漏，尤其是商机信息的泄漏」; existing solutions too generic for the industry 「现有解决方案过于通用，不够适合我们的行业/业务场景」; and lacking skills or resources to manage or deploy AI 「缺乏管理或部署 AI 所需要的技能或资源」. Administered in both studies.

**Objective AI knowledge.** Two multiple-choice items (0–2 correct): (1) "Which is a main application area of AI?"; 「以下哪个是人工智能的主要应用领域？」 (healthcare 「医疗健康」, financial services 「金融服务」, autonomous driving 「自动驾驶」, all of the above 「所有以上」; correct: all of the above); (2) "Which of the following is a machine-learning algorithm?"; 「以下哪个是人工智能/机器学习的算法？」 (linear regression 「线性回归」, decision tree 「决策树」, neural network 「神经网络」, all of the above 「以上都是」; correct: all of the above). Mean = 0.30, *SD* = 0.52; the two-item score is reported as a supplement, and the application-domain item provides the objective AI knowledge covariate.

***Study 2 measures***

**Phrase recognition.** Single item: "Which of the following was the phrase you set for your login welcome message at the start of the study?"; 「以下哪一条是你在研究开始时为登录欢迎语设置的短语？」 (own phrase presented among other participants' phrases); 86.7% and 80.0% across conditions. Developed for this study.

**Cue evocation.** Open-ended response to the prompt "When you saw this welcome message at login each day, what did it typically bring to mind?"; 「每天登录时看到这条欢迎语，你通常会想到什么？」 Coded 0–3 (0 = nothing/decorative … 3 = an explicit reminder that LLM output can err and should be verified); inter-rater $\kappa$ = .83.

**Cue–episode association.** Four items (three verification-related plus a reverse-scored decorative item), 1–5: (1) "The message brought the earlier self-explanation to mind"; 「这条消息让我想起了之前的自我解释环节。」 (2) "The message reminded me that AI output needs to be checked"; 「这条消息会提醒我需要核查 AI 的输出。」 (3) "The message was connected to what I had written during the training session"; 「这条消息和我在培训环节写下的想法有关联。」 (4, reverse-scored) "The message was just a decoration on the page, with no particular meaning"; 「这条消息只是页面上的装饰，没有特别含义。」 Cronbach's $\alpha$ = .81.

**Phrase attributes (balance check).** Five items, 1–5: "This phrase is personally meaningful to me" 「这条短语对我个人有特别的意义。」; "This phrase is visually noticeable on the page" 「这条短语在页面上很显眼。」; "This phrase is easy to remember" 「这条短语容易记住。」; "I like this phrase" 「我喜欢这条短语。」; "This phrase captured my attention when I logged in" 「登录时这条短语会吸引我的注意。」 Cronbach's $\alpha$ = .91.

The survey-based covariates and supplements appear above and were developed for this study; domain capability is computed from administrative sales records.

## Appendix D. Robustness Tests for Study 1 and Study 2

Appendix Table D1 consolidates the treatment-effect estimates across the alternative outcome codings examined in the two studies; the analyses behind each row are described in the study-specific robustness sections below. The estimates are also insensitive to the coding of the domain-capability covariate: recoding it as a binary indicator of any recorded sales leaves the treatment estimates unchanged (Study 1: b = 0.105, $p < .001$; Study 2 study day × retrieval cue: b = 0.008, p = .013). The same holds when objective AI knowledge is coded as the supplementary two-item score (0–2) or when the covariates are omitted altogether (Study 1: b = 0.105–0.106; Study 2: b = 0.008, p = .013 in every specification). Replacing domain capability with longer-horizon experience measures (years of sales experience, firm-specific sales tenure, or a past-year large-deal indicator) likewise leaves the estimates intact: Study 1 treatment estimates range from b = 0.103 to 0.110 (all $p < .001$), and the Study 2 interaction remains b = 0.008 (p = .013) throughout.

**Appendix Table D1. Treatment-Effect Estimates Across Alternative Outcome Codings in Studies 1 and 2**

| Outcome coding | Study 1: Self-explanation | Study 2: Study Day × Retrieval Cue |
|---|---|---|
| Primary error-detection rate | *b* = 0.105*** (0.012) | *b* = 0.008* (0.003) |
| High detection (median split, binary) | *OR* = 3.9*** | *OR* = 1.17*** |
| Most consequential error corrected (binary, email level) | | *OR* = 1.24*** |
| Most-consequential errors corrected (share per day) | | *b* = 0.021*** (0.005) |
| False-positive rate (correct content flagged) | *b* = −0.030*** (0.004) | *b* = 0.070 × $10^{-3}$ (0.001) |
| Net detection (detection − false positives) | *b* = 0.136*** (0.013) | *b* = 0.008* (0.003) |

***Note.*** The Study 1 column reports the self-explanation coefficient (odds ratio for binary codings) from covariate-adjusted models (*N* = 400). The Study 2 column reports the study day × retrieval-cue interaction (odds ratio for binary codings) from mixed-effects or generalized-estimating-equation models with the three covariates (*N* = 1,920 participant-days; the email-level model uses 5,760 email-level observations). Standard errors are in parentheses; odds ratios are reported with significance levels only. The false-negative rate is one minus the detection rate, so its estimates mirror the first row with reversed sign. * $p < .05$, *** $p < .001$ (two-tailed).

### *Robustness Tests for Study 1*

The regression estimates reported in Table 5 of the main text are algebraically equivalent to a one-way ANOVA and ANCOVA. The equivalent statistics are $F(1, 398) = 78.91$, $p < .001$, partial $\eta^2 = .17$ without covariates and $F(1, 395) = 77.43$, $p < .001$, partial $\eta^2 = .16$ with the three covariates, whose F tests are likewise nonsignificant (objective AI knowledge $F(1, 395) = 0.52$, p = .47; domain capability F = 0.24, p = .63; oversight orientation F = 0.18, p = .67). A net-detection measure (error detection minus false positives) yielded an even larger advantage (b = 0.136, $p < .001$). We tested the stability of the Study 1 findings and evaluated alternative explanations. First, we examined whether the results were sensitive to alternative operationalizations of verification behavior by focusing on false positives. Model-free comparisons showed a lower false-positive rate in the self-explanation condition (0.09 versus 0.12; Table 4), indicating that the error-detection advantage did not come at the cost of indiscriminate flagging. In line with this pattern, regression analyses controlling for objective AI knowledge, domain capability, and oversight orientation indicated that self-explanation reduced false-positive rates ($b = -0.030$, $p < .001$) rather than increasing them.

Second, we examined whether additional demographic controls altered the results. Including age and organizational tenure as covariates alongside the three focal covariates did not attenuate the self-explanation effect, $F(1, 393) = 76.12, p < .001$, *and neither age* ($p = .16$) *nor organizational tenure* ($p = .13$) was a significant predictor of error detection. Third, dichotomizing each participant's error-detection rate at the sample median and re-estimating the model as a logistic regression yields a large and significant self-explanation effect (odds ratio = 3.8 without adjustment and 3.9 with objective AI knowledge, domain capability, and oversight orientation as covariates; both $p < .001$), confirming that the result is not an artifact of the graded coding.

Together, these robustness analyses indicate that the Study 1 results are stable across alternative model specifications and sets of control variables, and that the observed effect of self-explanation reflects a robust behavioral difference.

### *Robustness Tests for Study 2*

We tested the stability of the Study 2 findings and evaluated alternative explanations for the observed trajectory effects. First, we examined whether the results depended on how error detection was operationalized and whether improvements reflected spurious vigilance rather than meaningful oversight. In addition to our primary measure of error-detection rate, we constructed a false positive rate capturing instances in which participants flagged or modified LLM-generated content that was in fact correct. Re-estimating the mixed-effects models using the false positive rate as the dependent variable revealed no significant retrieval-cue main effect ($b = -0.001$, $p = .85$) and no significant study day × retrieval-cue interaction ($b = 0.070 \times 10^{-3}$, $p = .93$). This pattern offers little evidence that the retrieval cue increased indiscriminate suspicion or overcorrection. Together with the positive net-detection result, this pattern is more consistent with sustained detection of genuine errors than with increased false alarms.

Second, we assessed whether the findings were sensitive to assumptions about within-participant dependence in repeated measures. Our main analyses used linear mixed-effects models with random intercepts for participants. As a robustness check, we re-estimated the models allowing the slope of study day to vary across participants. In this random-slope specification, the estimated decline in error detection over time ($b = -0.025$, $p < .001$) and the study day × retrieval-cue interaction ($b = 0.008$, $p = .013$) remained substantively unchanged relative to the random-intercept models in Table 7.

Third, we examined whether task-specific factors or fatigue effects could account for the observed erosion in error detection. We re-estimated the models using study-day fixed effects, allowing each day to have its own intercept. The study day × retrieval-cue interaction remained substantively unchanged ($b = 0.008$, $p = .013$), indicating that the trajectory differences are not driven by idiosyncratic properties of specific days or by general fatigue in later days. The interaction is also robust to disaggregation: estimated over the 5,760 email-level observations with participant random intercepts, the study day × retrieval-cue interaction remains positive and significant ($b = 0.007$, $p = .018$), and a logistic mixed-effects model estimated at the level of the 17,280 embedded errors yields the same conclusion (day × cue odds ratio = 1.03, $p = .035$).

Fourth, differential attrition cannot account for the results: every participant in the analytic sample completed all eight study days and all 24 embedded test emails, so the estimation sample is complete by construction and no complete-case restriction is required.

Fifth, we examined whether the retrieval cue merely increased general vigilance rather than altering the trajectory of behavior. Because Day 0 preceded cue generation and display, the Day 0 difference between conditions is a random baseline difference (Mean = 0.640 versus 0.622; $t = 0.71$, $p = .48$). Re-estimating the model on Days 1–7 with Day 0 performance included as a baseline covariate yields a positive study day × retrieval-cue interaction of similar magnitude that falls just short of conventional significance ($b = 0.007$, $p = .064$). The adjusted cue term is likewise not significant ($b = 0.023$, $p = .23$). This pattern is consistent with the interpretation that retrieval cues sustain access to prior reasoning over time instead of producing a one-time boost in attention.

Finally, the conclusions are robust to alternative outcome codings, as noted in the main text. Using the share of most-consequential errors corrected per day, error detection declines more steeply ($b = -0.101$, $p < .001$) and the interaction remains positive and significant ($b = 0.021$, $p < .001$), and the email-level binary indicator defined in the main text shows the same pattern in a logistic mixed-effects model (day × cue odds ratio = 1.24, $p < .001$); the cue-condition difference at Day 0 on this measure again reflects a pre-treatment baseline difference. Dichotomizing daily detection at the sample median and estimating a population-averaged logistic model (generalized estimating equations with exchangeable within-participant correlation) yields declining odds of high detection across days (odds ratio = 0.67, $p < .001$) and a positive study day × retrieval-cue interaction (odds ratio = 1.17, $p < .001$). The daily-rate models use the same analytic sample (1,920 participant-day observations from 240 participants); the email-level logistic model uses the corresponding 5,760 email-level observations from the same 240 participants.

Taken together, these analyses show that the decline in error detection is consistent across specifications and that the estimated cue attenuation is positive and similar in magnitude across models. The cue interaction is statistically significant in the random-slope, day-fixed-effects, email-level, error-level, net-detection, and most-consequential-error models, whereas the Day 0-adjusted Days 1–7 model yields a positive estimate of similar magnitude with lower precision ($b = 0.007$, $p = .064$).

To evaluate whether cue content alone explains the trajectory advantage, we recoded the 120 self-generated cues by form and instruction compliance (Table B1). Fifty-two cues (43.3%) were idiosyncratic personal tags, 65 (54.2%) were instruction-compliant cues with ordinary meaning, and only three (2.5%) were general fallibility statements that departed from the cue instruction. Excluding these three participants, the day × retrieval-cue interaction remained positive and significant ($b = 0.007$, $p = .016$; 237 participants and 1,896 participant-day observations). In the three-group model, the estimated daily slopes were −0.025 for the non-cue condition, −0.019 for idiosyncratic cues, and −0.016 for instruction-compliant cues with ordinary meaning (Figure D1). The idiosyncratic subgroup showed a flatter estimated trajectory than control ($b = 0.006$, $p = .159$), and its slope did not differ detectably from that of cues with ordinary meaning ($b = -0.003$, $p = .434$). The treatment effect survives removal of the three noncompliant generic statements, and ordinary-meaning cues show no detectable slope advantage over idiosyncratic cues. These results weigh against a strong content-only account. Because cue form was participant-selected, the subgroup comparisons are exploratory.

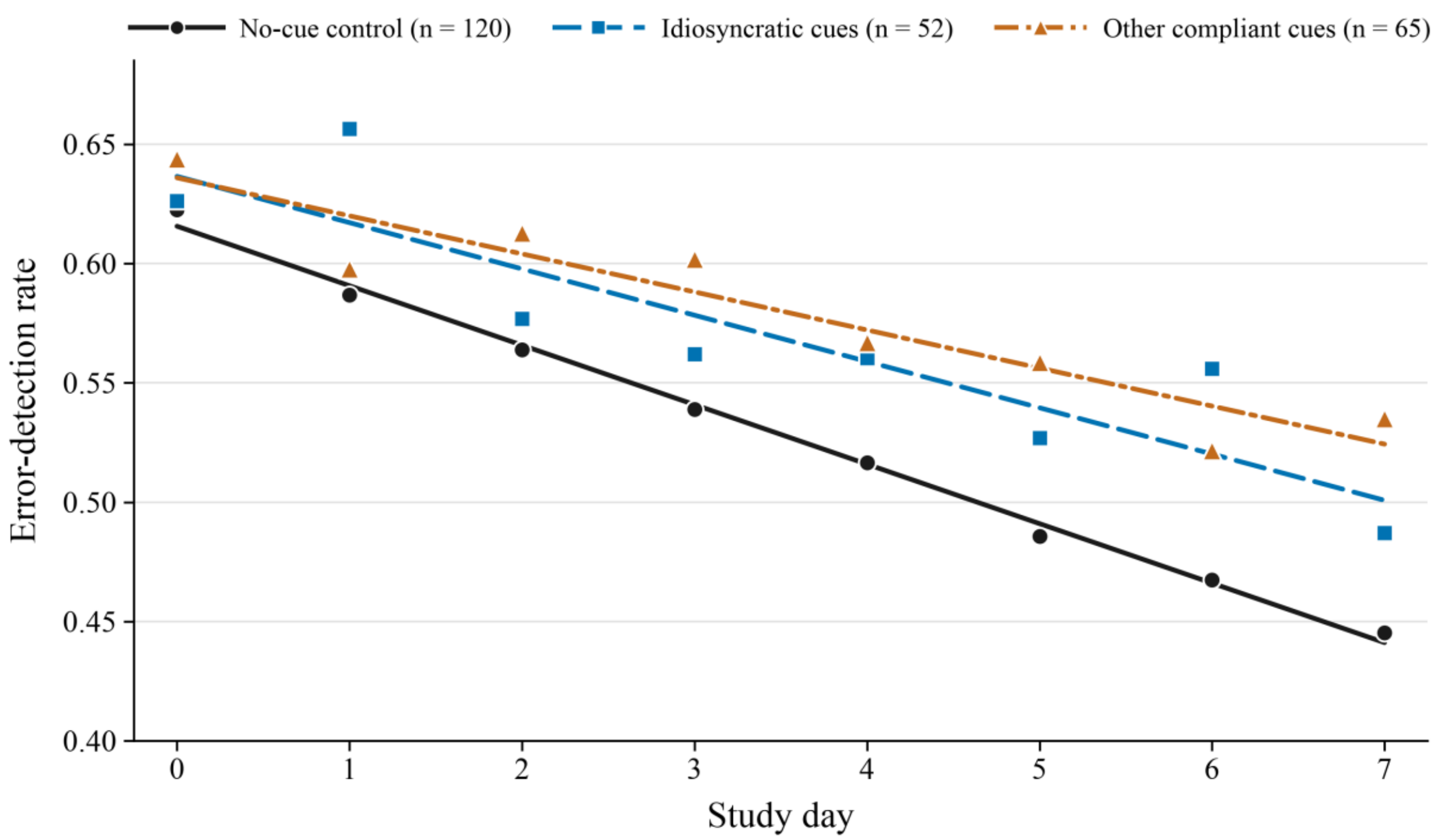

**Appendix Figure D1. Error-Detection Trajectories by Cue Form**

*Note.* Points show observed daily means; lines show covariate-adjusted linear trajectories from the mixed-effects model reported in the text. The analysis includes the non-cue condition (n = 120), idiosyncratic personal tags (n = 52), and other instruction-compliant cues (n = 65); the three cues that departed from the generation instruction are excluded.

Finally, we examined the mechanism-related moderation analyses reported in the main text's Study 2 results. The day × attribute-composite interaction was not significant in the full sample ($b = -0.001$, $p = .65$) or within the retrieval-cue condition (p = .35), and person-level error-detection slopes were unrelated to the attribute composite (r = −.03, p = .66), indicating that message salience, memorability, and liking do not shape trajectories. Coded evocation ratings predicted flatter trajectories across the sample (day × evocation b = 0.006, p < .001; person-level slope correlation r = .23, p < .001), tracking the randomized contrast. Because evocation is unevenly distributed across conditions (77.5% of cue-condition versus 48.3% of non-cue participants rated ≥ 2 on the 0–3 scale) and was measured once at the end of the study, within-condition dose-response tests are underpowered and were not significant (retrieval-cue condition day × evocation *p* = *.73*). We therefore interpret the randomized intervention effect, cue-evocation and cue–episode association measures, and message-attribute checks as convergent process evidence rather than as individual-level mediation or standalone identification of the retrieval mechanism.

## Appendix E. Correlation Matrices

Appendix E reports the correlation matrices for the Study 1 participant-level variables and the Study 2 participant-day variables. These correlations are descriptive; they are not used for hypothesis testing and do not carry mechanism claims.

**Appendix Table E1. Correlation Matrix for Study 1 Variables (*N* = 400)**

| Variable | (1) | (2) | (3) | (4) | (5) | (6) | (7) | (8) | (9) | (10) | (11) |
|---|---|---|---|---|---|---|---|---|---|---|---|
| (1) Error-detection rate | 1.00 | | | | | | | | | | |
| (2) False positive rate | −.16 | 1.00 | | | | | | | | | |
| (3) Verification time | .28 | −.26 | 1.00 | | | | | | | | |
| (4) Reference clicks | .17 | −.16 | .24 | 1.00 | | | | | | | |
| (5) Modification extent | .22 | −.21 | .33 | .23 | 1.00 | | | | | | |
| (6) Objective AI knowledge | .02 | .00 | −.05 | .03 | .03 | 1.00 | | | | | |
| (7) Domain capability (log sales) | −.06 | −.05 | −.11 | −.07 | −.08 | −.07 | 1.00 | | | | |
| (8) Oversight orientation | .03 | .04 | −.04 | .01 | −.04 | .02 | −.02 | 1.00 | | | |
| (9) Age | .04 | .02 | .03 | .02 | −.02 | −.01 | .04 | .02 | 1.00 | | |
| (10) Organizational tenure | −.05 | .05 | −.06 | −.04 | −.02 | −.03 | .06 | .03 | .56 | 1.00 | |
| (11) Gender (1 = female) | .06 | −.02 | −.08 | .00 | −.04 | .00 | −.06 | .05 | .03 | .06 | 1.00 |

***Note.*** Pearson correlations. Variables (1)–(5) are participant-level aggregates of the behavioral outcomes across the 10 tasks; (6)–(8) are participant-level covariates; (9)–(11) are demographic characteristics (gender coded 1 = female, 0 = otherwise). Lower triangle shown.

**Appendix Table E2. Correlation Matrix for Study 2 Day-Level Variables (*N* = 1,920 observations from 240 participants)**

| Variable | (1) | (2) | (3) | (4) | (5) | (6) | (7) | (8) | (9) | (10) | (11) |
|---|---|---|---|---|---|---|---|---|---|---|---|
| (1) Error-detection rate | 1.00 | | | | | | | | | | |
| (2) False positive rate | −.16 | 1.00 | | | | | | | | | |
| (3) Verification time | .28 | −.15 | 1.00 | | | | | | | | |
| (4) Reference clicks | .26 | −.15 | .57 | 1.00 | | | | | | | |
| (5) Modification extent | .27 | −.13 | .55 | .55 | 1.00 | | | | | | |
| (6) Objective AI knowledge | −.02 | .02 | .05 | .00 | .03 | 1.00 | | | | | |
| (7) Domain capability (log sales) | −.05 | .04 | .04 | −.03 | −.02 | .10 | 1.00 | | | | |
| (8) Oversight orientation | −.01 | .05 | .01 | −.02 | −.04 | .02 | .04 | 1.00 | | | |
| (9) Age | .03 | −.04 | .03 | .03 | .02 | −.01 | .04 | .02 | 1.00 | | |
| (10) Organizational tenure | .06 | .02 | .00 | .00 | .00 | .00 | .11 | .02 | .62 | 1.00 | |
| (11) Gender (1 = female) | −.02 | −.06 | −.03 | −.02 | −.01 | −.06 | .01 | −.09 | −.08 | −.08 | 1.00 |

***Note.*** Pearson correlations at the participant-day level. Variables (1)–(5) are daily behavioral outcomes; (6)–(8) are participant-level covariates; (9)–(11) are demographic characteristics (gender coded 1 = female, 0 = otherwise). Lower triangle shown.

## Appendix References

*Note*. This list reproduces, with identical numbering, the main-text references cited in the online appendices.